\documentclass{aastex631}

\begin{document}

\title{SN 2021pfs: A Type Ia Supernova Likely Affected by Progenitor Metallicity, as Revealed by Comparison with Its Twin Counterpart}

\author[0009-0003-9229-9942]{Abdusamatjan Iskandar} 
\affiliation{Xinjiang Astronomical Observatory, Chinese Academy of Sciences, Urumqi, Xinjiang, 830011, China}
\affiliation{School of Astronomy and Space Science, University of Chinese Academy of Sciences, Beijing 100049, China}

\author[0000-0002-7334-2357]{Xiaofeng Wang} 
\correspondingauthor{Xiaofeng Wang}
\email{wang\_xf@mail.tsinghua.edu.cn}
\affiliation{Physics Department, Tsinghua University, Beijing 100084, People’s Republic of China}

\author{Ali Esamdin} 
\correspondingauthor{Ali Esamdin}
\email{aliyi@xao.ac.cn}
\affiliation{Xinjiang Astronomical Observatory, Chinese Academy of Sciences, Urumqi, Xinjiang, 830011, China}
\affiliation{School of Astronomy and Space Science, University of Chinese Academy of Sciences, Beijing 100049,  China}

\author{Wenxiong Li} 
\affiliation{National Astronomical Observatories, Chinese Academy of Sciences, 100101, China}

\author{Xiangyun Zeng} 
\affiliation{Center for Astronomy and Space Sciences, China Three Gorges University, YiChang, 443000, People's Republic of China}
\affiliation{College of Mathematics and Physics, China Three Gorges University, YiChang, 443000, China}

\author{Ruifeng Huang} 
\affiliation{Physics Department, Tsinghua University, Beijing 100084, People’s Republic of China}
\author{Maokai Hu} 
\affiliation{Physics Department, Tsinghua University, Beijing 100084, People’s Republic of China}

\author{Guoliang L$\ddot{\rm u}$} 
\affiliation{School of Physical Science and Technology, Xinjiang University, Urumqi 830046, People’s Republic of China}

\author{D. Andrew Howell} 
\affiliation{Las Cumbres Observatory, 6740 Cortona Drive Suite 102, Goleta, CA 93117-5575, USA}
\affiliation{Department of Physics, University of California, Santa Barbara, CA 93106-9530, USA}

\author{Curtis McCully} 
\affiliation{Las Cumbres Observatory, 6740 Cortona Drive Suite 102, Goleta, CA 93117-5575, USA}

\author[0000-0003-2732-4956]{Samuel Wyatt} 
\affiliation{Astrophysics Science Division, NASA Goddard Space Flight Center, Greenbelt, MD 20771, USA}

\author[0000-0002-1125-9187]{Daichi Hiramatsu} 
\affiliation{Department of Astronomy, University of Florida, 211 Bryant Space Science Center, Gainesville, FL 32611-2055 USA}
\affiliation{Center for Astrophysics \textbar{} Harvard $\&$ Smithsonian, 60 Garden Street, Cambridge, MA 02138-1516, USA} 
\affiliation{The NSF AI Institute for Artificial Intelligence and Fundamental Interactions, Cambridge MA 02139, USA}

\author{Estefania Padilla Gonzalez} 
\affiliation{Department of Physics and Astronomy, The Johns Hopkins University, Baltimore MD 21218, USA}

\author{Craig Pellegrino} 
\affiliation{Goddard Space Flight Center, Greenbelt, MD 20771, USA}

\author{Megan Newsome} 
\affiliation{Department of Astronomy, University of Texas, Austin, TX, 78712,USA}

\author[0000-0002-1296-6887]{Llu\'is Galbany} 
\affiliation{Institute of Space Sciences (ICE-CSIC), Campus UAB, Carrer de Can Magrans, s/n, E-08193 Barcelona, Spain}
\affiliation{Institut d'Estudis Espacials de Catalunya (IEEC), Castelldefels (Barcelona)  08860, Spain}
\author[0000-0003-4102-380X]{David J. Sand} 
\affiliation{Steward Observatory, University of Arizona, Tucson AZ 85721-0065, USA}

\author[0000-0001-5510-2424]{Nathan Smith}
\affiliation{Steward Observatory, University of Arizona, Tucson AZ 85721-0065, USA}




\author{Huei Sears} 
\affiliation{Department of Physics and Astronomy, Rutgers, the State University of New Jersey, Piscataway, NJ 08854-8019, USA}

\def\dm15{$1.13\pm0.06~$}
\def\ABmag{$-19.28\pm0.40~$}
\def\mbmax{$~13.82~\pm~0.03~$}
\def\tbmax{$~59391.49~\pm~0.34~$}
\def\DM{$~32.63~\pm~0.39~$}
\def\BVmax{$~0.05~\pm~0.05~$}
\def\ebvhost{$~0.117~\pm~0.060~$}
\def\nimass{$0.48 \pm 0.05$}
\def\Lbolo{($1.00 \pm 0.14)~\times~10^{43}~$}
\def\t0{$~59373.22~\pm~0.06~$}
\def\tbrise{$18.27~\pm~0.35~$}
\def\sbv{$~0.90~\pm~0.03~$}

\newcommand{\abu}[1]{\textcolor{red!50!blue}{\textbf{Abu:} }{\textcolor{blue}{#1}}}


\begin{abstract}
We present extensive photometric and spectroscopic observations of the normal type Ia supernovae (SNe Ia) 2021pfs, which occurred in the Seyfert 2 galaxy NGC 5427 at a redshift 0.009. SN 2021pfs reached an absolute \textit{B}-band peak magnitude of $M_{\rm max}(B)=$\ABmag mag and a post-peak decline rate of $\Delta m_{15}(B)=$\dm15mag. The observed properties of this nearby SN Ia closely resemble those of SN 2011fe, including the main optical spectroscopic features and photometric evolution. Despite their similar decline rates, SN 2021pfs rose more rapidly in the \textit{U} band but more slowly in the \textit{r} and \textit{i} bands compared to SN 2011fe in very early phases. This photometric difference, particularly at short wavelengths, can introduce a systematic uncertainty of up to  $\sim$12\% in distance estimates. Analysis of the host galaxy's local and global environment shows an environment consistent with producing a higher-metallicity progenitor for SN 2021pfs than that of SN 2011fe.
This higher progenitor metallicity may explain the observed photometric discrepancy and the resulting distance between SN 2021pfs and SN 2011fe, though a larger sample of such "twin" SNe Ia is needed to confirm this trend and assess its impact on cosmological measurements.  

\end{abstract}



\section{Introduction} \label{sec:intro}
Type Ia supernovae (SNe Ia) have a well-known ``standard candle” behavior, especially the tight correlation between their peak luminosity and light-curve width (or color-stretch) \citep{Phillips1993ApJ,Riess1996ApJ, Phillips1999AJ,Wang2005ApJ,Guy2007A&A...466...11G}. This correlation (dubbed the $Phillips$ relation) has established SNe Ia as crucial tools probing the expansion history of the Universe \citep{Riess1998AJ,Perlmutter1999ApJ...517..565P,Perlmutter1999PhRvL..83..670P,Riess1996ApJ,Burns2018ApJ,Guy2007A&A...466...11G}. 
SNe Ia are generally believed to arise from the thermonuclear explosions of carbon-oxygen white dwarfs (CO WDs) in binary systems. However, their progenitor systems and explosion mechanisms \citep{Hoyle1960ApJ...132..565H,Woosley1986ApJ,Nugent2011Natur,Wang2012NewAR,Bloom2012,Darnley2014A&A,Maoz2014,Jha2019NatAs} remain poorly understood. In the single-degenerate (SD) scenario, the companion star could be a main-sequence, red giant, or helium star. The CO WD accretes material from the companion star until it nears the Chandrasekhar mass, triggering an SN Ia explosion \citep{Chandrasekhar1957,Whelan1973,Nomoto1982,Han2004MNRAS.350.1301H,Wang2009MNRAS}. In the double-degenerate (DD) scenario, two WDs merge, leading to an explosion \citep{Iben1984,Webbink1984,Pakmor2012}. A core-degenerate (CD) scenario has also been proposed, wherein a CO WD merges with the core of an asymptotic giant branch (AGB) star, exploding as a SN Ia after a delayed evolution \citep{Kashi2011MNRAS.417.1466K, Soker2013IAUS..281...72S, Soker2014MNRAS.437L..66S}. 

Various explosion models have been studied, including pure deflagration, supersonic detonation, delayed detonation (deflagration-to-detonation transitions), surface helium shell detonations, and even detonations triggered by direct WD collisions \citep{Nomoto1984ApJ...286..644N,Hoeflich1995ApJ...444..831H,H2002ApJ...568..779H,H2006NuPhA.777..579H,Thielemann1986A&A...158...17T,Shen2018ApJ...854...52S,Polin2019ApJ...873...84P,Rosswog2009ApJ...705L.128R,Pakmor2010Natur.463...61P,Pakmor2012}.
The double-detonation scenario posits that a sub-Chandrasekhar mass CO WD accretes a helium layer on its surface; a detonation of this helium shell can trigger a thermonuclear explosion in the core before the Chandrasekhar mass limit is reached \citep{Fink2007A&A...476.1133F,Sim2010ApJ...714L..52S,Shen2018ApJ...854...52S}. The helium fuel may come from a non-degenerate companion star or a lower-mass helium WD. Numerous simulations have been conducted to explore this model \citep{Fink2007A&A...476.1133F,Sim2010ApJ...714L..52S,Shen2018ApJ...854...52S,Polin2019ApJ...873...84P}. For example, \cite{Sim2010ApJ...714L..52S} found that for CO WD with masses ranging from 0.97 to 1.15 $\rm M_\odot$, the synthesized nickel mass could be 0.3-0.8 $\rm M_\odot$. The double-detonation model is one of the main proposed mechanisms for SN Ia explosions and is often invoked to explain early-time flux excess in some SNe Ia \citep{Shen2018ApJ...854...52S,Polin2019ApJ...873...84P}.

The luminosity at maximum light, light-curve width, and photospheric temperature of SNe Ia mainly depend on the mass of the radioactive element $^{56}$Ni produced in the explosion \citep{Phillips1993ApJ,Kasen2007ApJ...656..661K}. The  \textit{Phillips relation} 
enabled the use of SNe Ia in precision cosmology. However, a growing body of evidence reveals the diversity of SNe Ia in both photometry and spectroscopy. Several peculiar subclasses have been identified, including over-luminous 1991T-like \citep{Filippenko1992ApJ}, super-Chandrasekhar 2003fg-like \citep{Howell2006Natur.443..308H}, subluminous  1991bg-like \citep{Filippenko1992AJ....104.1543F}, 2002cx-like \citep{Li2003PASP..115..453L}, and 2002es-like \citep{Ganeshalingam2012ApJes} events. Even among ``normal” SNe Ia, one can distinguish subtypes by spectroscopic differences.  
For instance, \cite{Branch2006PASP} defined four spectroscopic subclasses: Core-Nomal (CN), Shallow-Silicon (SS), Broad-Line (BL), and Cool (CL)-based on the pseudo-equivalent width of Si {\sc ii} $\lambda$5972 and $\lambda$6355 \citep{Burrow2020ApJ...901..154B}. 
\cite{Benetti2005ApJ} introduced a classification by Si {\sc ii} velocity gradient: high-velocity gradient (HVG), low-velocity gradient (LVG), and FAINT subclasses. \cite{Wang2009HVNV} divided normal SNe Ia into normal-velocity (NV) and high-velocity (HV) subclasses based on the velocity of Si {\sc ii} $\lambda$6355 near maximum light. These subclasses exhibit different observational characteristics, suggesting that they may originate from different progenitor systems or explosion mechanisms. 


Connections between SN Ia observables and host galaxy properties have been extensively studied \citep{Nugent1995ApJ,Timmes2003ApJ...590L..83T,Benetti2004MNRAS,Mazzali2006MNRAS.369L..19M,Foley2013ApJ...769L...1F,Wang2013Sci...340..170W,Pan2015MNRAS.446..354P,Brown2020ApJ...890...45B,Burgaz2026A&A...705A..76B}. For example, \cite{Lentz2001ApJ...547..402L} found that SNe Ia with higher progenitors metallicity yields larger blueshifts in Si {\sc ii} absorption lines. Observationally, \cite{Foley2013ApJ...769L...1F} suggested that  variations in the UV continuum among SNe Ia could be linked to progenitor metallicity. SNe Ia with higher Si {\sc ii} $\lambda$6355 velocities tend to occur in more massive galaxies and closer to their inner regions \citep{Wang2009HVNV,Wang2013Sci...340..170W,Pan2015MNRAS.446..354P}, and SNe Ia with stronger Ca {\sc ii} near-infrared (NIR) triplet absorption lines prefer low-mass galaxies with higher star formation rates \citep{Pan2015MNRAS.446..354P}. \cite{Brown2020ApJ...890...45B} found no clear correlation between the UV–optical colors of SNe Ia and the mass or metallicity of their host galaxies. Nevertheless, for SNe Ia with similar light-curve widths, higher progenitor metallicity is expected to lead to lower luminosities and smaller yields of $^{56}$Ni \citep{Timmes2003ApJ...590L..83T,Mazzali2006MNRAS.369L..19M,Brown2020ApJ...890...45B}. This effect could partly account for the observed luminosity differences between otherwise ``twin” SNe Ia.

The metallicity of the progenitor can influence the nucleosynthesis yields, particularly the mass of Ni synthesized during the explosion, which in turn directly affects the peak brightness and light curve shape \citep{Timmes2003ApJ...590L..83T,Podsiadlowski2006astro.ph..8324P}. Preivous studies found variations in metallicity can lead to significant differences in Ni production \citep{Timmes2003ApJ...590L..83T}. Even after light-curve shape corrections, SNe Ia in massive, passive host galaxies remain brighter (approximately 3$\sigma$) than those in low-mass, star-forming host galaxies \citep{Kelly2010ApJ...715..743K,Lampeitl2010ApJ...722..566L,Sullivan2010MNRAS.406..782S}. Although this systematic effect might not be that significant, it is real and suggests that current luminosity correction methods do not fully account for the physical diversity of SNe Ia, with intrinsic scatter still present.
Several studies have reported a correlation between host metallicity and Hubble 
residuals, suggesting that metallicity-driven variations in peak brightness could introduce systematic biases in distance measurements \citep{Gallagher2008ApJ...685..752G,Rigault2013A&A...560A..66R,Childress2013ApJ...770..108C,Moreno-Raya2018MNRAS.476..307M}. More recent analyses involving large samples have continued to explore such an effect, finding evidence that metallicity corrections may improve the precision of cosmological parameter constraints \citep{Brout2022ApJ...938..110B,Dixon2025MNRAS.538..782D}.

In this paper, we present extensive observations of the normal SN Ia 2021pfs and compare them in detail to SN 2011fe. In Section \ref{sec:obs} we describe the optical observations and data reduction. In Section \ref{sec:opl} we analyze the multi-band light and color curves of SN 2021pfs, and determine its time of first light. Section \ref{sec:sepc} examines the spectral evolution. Section \ref{sec:discuss} discussed the quasi-bolometric light curve, host galaxy properties, and implications for the explosion physics. We summarize our findings in Section \ref{sec:con}. In the Appendix, the photometric local reference stars in the SN 2021pfs field are listed in Table \ref{tab_ref}, and Table \ref{tab:lco_lc} lists the photometry of SN 2021pfs obtained from LCO.

\section{OBSERVATIONS AND DATA REDUCTION} 
\label{sec:obs}
\begin{figure}[ht!]
    \centering
	\includegraphics[width=4.5in]{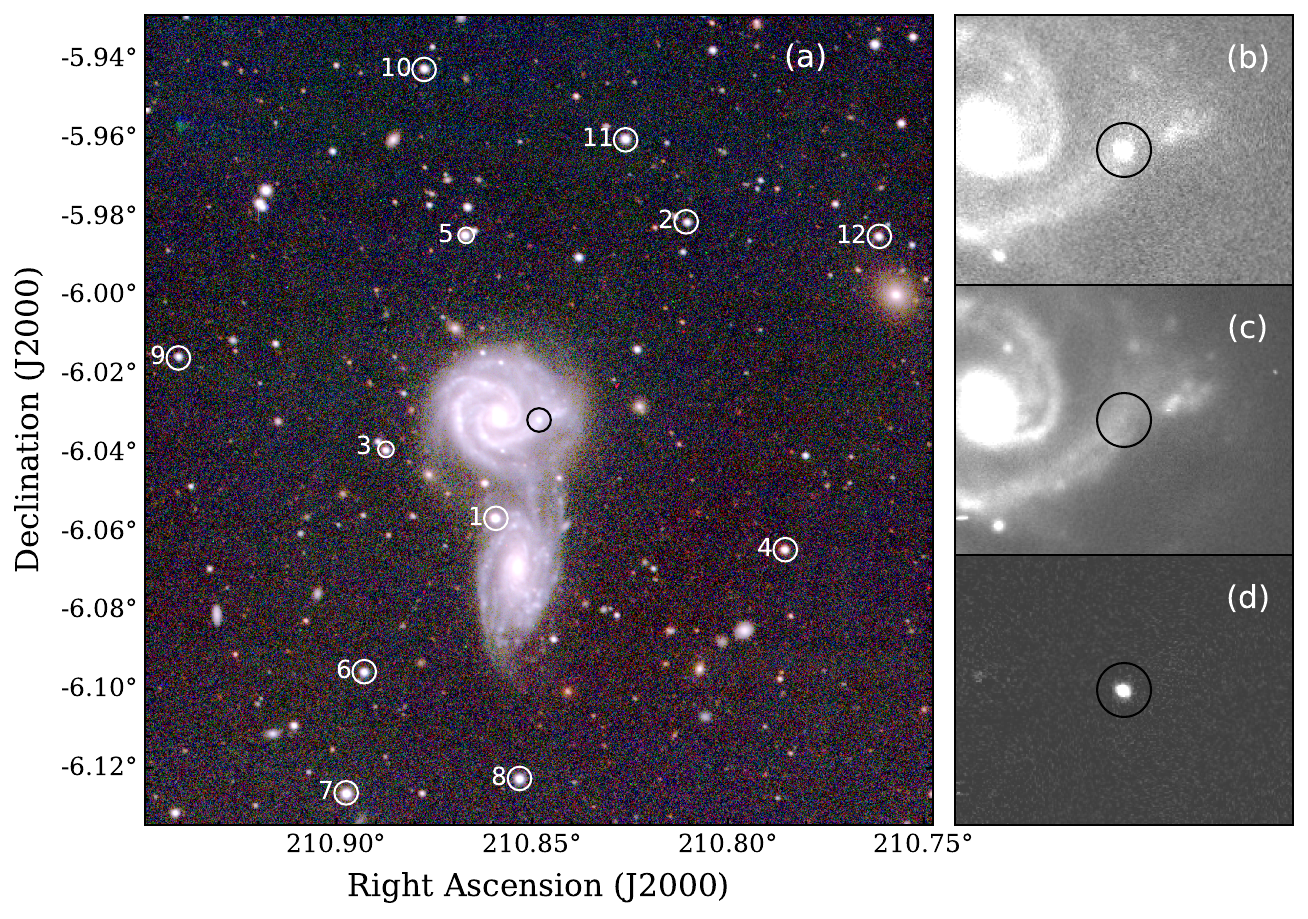}
	\caption{(a): An \textit{RGB} composite image taken with the LCO 1-m telescope (CPT) on 2021 June 12, constructed using the \textbf{APLpy} package \citep{Robitaille2012asclsoft08017R} from the \textit{rgB}-band images. The black circle indicates SN 2021pfs, and the numbers label reference stars from the APASS catalog. Reference stars are marked with white circles. (b): LCO 1-m \textit{g}-band image on 2021 June 12.74 UT, showing a 0.01-degree field centered on SN 2021pfs. (c) LCO 1-m \textit{g}-band template image on 2022 June 04.79 UT. (d) Difference image resulting from the subtraction of panel (c) from (b) using HOTPANTS. Panels (c) and (d) have the same scale and field of view as (b).}
	\label{Fig:pfs_5427}
\end{figure}

SN 2021pfs was first detected on 2021 June 09 (UT; MJD 59374.218) with \textit{g}-band magnitude of $g$ $\approx$ 19.42 mag by the Zwicky Transient Facility (ZTF) Palomar \citep{Munoz-Arancibia2021TNSTR,Bellm2019PASP..131f8003B}. The last nondetection was two days earlier  (MJD 59372.310) with a limiting magnitude of \textit{r} $\approx$ 16.87 mag~\citep{Munoz-Arancibia2021TNSTR}. SN 2021pfs lies approximately 36.3$\arcsec$ west and 3.8$\arcsec$ south of the center of its host galaxy, NGC 5427 (Figure \ref{Fig:pfs_5427}(a)). 
Follow-up spectroscopy on MJD 59375.20 indicates that SN 2021pfs was a very young, normal SN Ia \citep{Wyatt2021TNSCR2003}, at $\approx-16.3$ days before the maximum luminosity according to the spectroscopic classification with SNID~\citep{Blondin2007ApJ...666.1024B}. The host galaxy NGC 5427~\citep{Wyatt2021TNSCR2003} has a redshift of 0.008733~\citep{Theureau1998A&AS}.

Through the Global Supernova Project, we obtained a series of \textit{uUBgVrizY} images using the Las Cumbres Observatory (LCO) global telescope network~\citep{Brown2013LCO}, covering from about -17 days to +80 days relative to the \textit{B}-band maximum light. Image reduction  was performed with the LCO \textbf{BANZAI} pipeline~\citep{Valenti2016MNRAS} and a Python-based pipeline. Image subtraction was carried out using the HOTPANTS package~\citep{Becker2015ascl} (Figure \ref{Fig:pfs_5427}(b)-(d)). Instrumental magnitudes were measured with SExtractor \citep{Bertin1996}. Photometric calibration was achieved using APASS catalog stars for \textit{BgVri} bands \citep{Hende2016yCat} and Sloan Digital Sky Survey (SDSS) catalog \citep{Zheng2017ApJ...841...64Z} for \textit{u} band. 
The finder chart including the SN and comparison stars is shown in Figure \ref{Fig:pfs_5427}(a). The coordinates and the $UBVgri-$band magnitudes of the reference stars are tabulated in Table \ref{tab_ref}.
Instrumental \textit{U}-band magnitudes were transformed to the standard Johnson system using color terms derived from observations of Landolt standard stars~\citep{Clem2016AJ} with the Nanshan One-meter Wide-field Telescope (NOWT)~\citep{Bai2020RAA}. The \textit{z}- and \textit{Y}-band photometry were calibrated against the Panoramic Survey Telescope and Rapid Response System (Pan-STARRS1) catalog~\citep{Chambers2016arXiv161205560C}. 
 
We obtained a total of 12 low-resolution optical spectra with phases ranging from $\sim -16$ days to $\sim$ +54 days relative to the maximum light. 
The optical spectra were mainly obtained from the LCO-2m/FLOYDS~\cite[through the Global Supernova Project]{Brown2013LCO}.
The journal of spectroscopic observations is presented in Table \ref{tab:specjo}. Spectral reductions were performed using standard tasks in IRAF\footnote{Image Reduction and Analysis Facility; distributed by NOAO, operated by AURA under cooperative agreement with the National Science Foundation}, including bias subtraction, extraction, wavelength calibration, and flux calibration with spectrophotometric standard stars. We then applied corrections for atmospheric extinction and telluric absorption lines. Two archival early spectra were retrieved from the Transient Name Server (TNS), taken with the SEDM on the 60-inch Palomar telescope (ZTF P60), and the Binospec spectrograph on the 6.5m Multiple Mirror Telescope (MMT). 

\startlongtable
\begin{deluxetable*}{cccccccc}
\tablecolumns{3} 
\tablewidth{0pc} 
\tabletypesize{\scriptsize}
\tablecaption{Spectroscopic Observations of SN 2021pfs}
\tablehead{\colhead{Date (UTC)} &\colhead{MJD} &\colhead{Phase$^a$ (days)}  &\colhead{Telescope} &\colhead{Instrument} &\colhead{Exposure (s)} &\colhead{Range ($\mathring{\rm A}$)}              &\colhead{$\Delta \lambda$ ($\mathring{\rm A}$)}}
\startdata
\centering
    2021-06-10T04:45:43	&59375.20 	&-16.3 	&MMT6.5m   	    &Binospec &7200	&3902.8- 9134.1	&1.3\\
    2021-06-10T07:27:27	&59375.31 	&-16.2 	&OGG-2m	        &FLOYDS	  &3600	&3500.4-10000.2	&2.3\\
    2021-06-12T08:22:22	&59377.35 	&-14.2 	&OGG-2m	        &FLOYDS	  &2700	&3499.4-10001.1	&2.3\\
    2021-06-13T07:37:34	&59378.32 	&-13.2 	&ZTF-P60	    &SEDM     &2160	&3776.7- 9223.3	&25.5\\
    2021-06-14T07:34:48	&59379.32 	&-12.2 	&OGG-2m	        &FLOYDS	  &2700	&3499.9-10001.1	&2.3\\
    2021-06-17T06:43:55	&59382.28 	&-9.2 	    &OGG-2m	        &FLOYDS	  &2700	&3499.9-10001.1	&2.3\\
    2021-06-23T07:39:16	&59388.32 	&-3.2 	    &OGG-2m	        &FLOYDS	  &2700	&3500.4- 9999.5	&2.3\\
    2021-06-27T12:13:18	&59392.51 	&1.0 	    &COJ-2m	        &FLOYDS	  &900	&3501.0-10000.2	&2.3\\
    2021-07-03T12:03:29	&59398.50 	&7.0 	    &COJ-2m	        &FLOYDS	  &900	&3800.1- 9999.3	&2.3\\
    2021-07-07T11:26:23	&59402.48 	&11.0 	    &COJ-2m	        &FLOYDS	  &900	&3499.2-10000.1	&2.3\\
    2021-08-03T05:56:34	&59429.25 	&37.8 	    &OGG-2m	        &FLOYDS	  &899	&3500.4- 9999.1	&2.3\\
    2021-08-19T08:36:05	&59445.36 	&53.9 	    &COJ-2m	        &FLOYDS	  &1200	&3500.3-10000.2	&2.3\\ 
\enddata
\tablenotetext{a}{Relative to the date of \textit{B}-band maximum brightness (MJD 59391.49).} 
\label{tab:specjo}
\end{deluxetable*}

\section{Photometry}
\label{sec:opl}
\subsection{Multiband Light Curves, Color Curves, and Distance} 
\begin{figure}[ht!]
     \centering
 	\includegraphics[width=6.in]{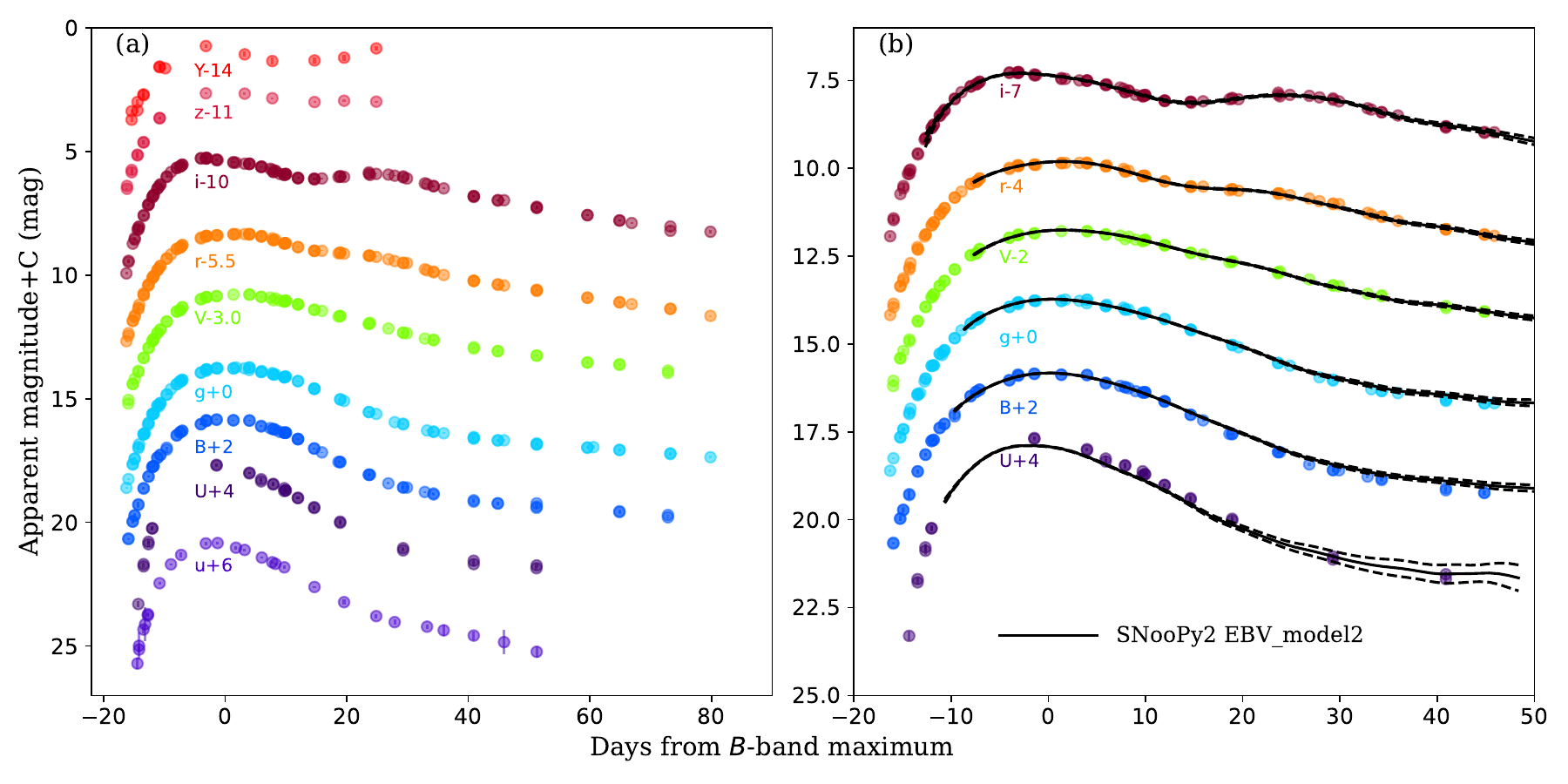}
 	\caption{(a): The multiband light curves of SN 2021pfs, obtained using facilities from LCO. (b): Optimal fitting light curves from \textbf{SNooPy2} (solid lines). The dashed black lines indicate the $1-\sigma$ uncertainty range associated with the best-fit light curve templates. The light curves have been vertically shifted across all panels. 
 }
 	\label{fig:allbandlc}
 \end{figure}

The multi-band light curves of SN 2021pfs are displayed in Figure \ref{fig:allbandlc}(a), and the photometry is given in Table \ref{tab:lco_lc}. 
By fitting low-order polynomials to the near-peak LCO light curves, we find that SN 2021pfs reached \textit{B}-band maximum  at \mbmax mag on MJD \tbmax, and \textit{V}-band maximum at 13.77 $\pm$ 0.02 mag on MJD 59391.87 $\pm$ 0.24. The light-curve shape is typical of a normal SN Ia, showing a shoulder in the \textit{r} band and a prominent secondary maximum in the \textit{i} band, which becomes more pronounced in the \textit{Y} band. The peak magnitudes and dates for each band are listed in Table \ref{tab:phot_par}, which shows that the rise times are shorter in the bluer filters and longer in the red ones.

We fit the multi-color light curves of SN 2021pfs using the SuperNovae in object-oriented Python (SNooPy2) package~\citep{Burns2011AJ}. Figure \ref{fig:allbandlc}(b) shows the best-fit template light curves with EBV\_model2 (solid lines). 
From the fits, we determine a post-peak decline of $\Delta m_{15}(B)$ = \dm15~mag, defined as the decline in the \textit{B}-band magnitude over 15~days after maximum \citep{Phillips1999AJ}, and a color-stretch parameter of $s_{BV}$ = \sbv \citep{Burns2011AJ}, both of which are close to the normal values for normal SNe Ia. The distance modulus derived from the SNoopy2 fit using the EBV$\_$model2 is $32.65 \pm 0.09$ mag. The decline rate of SN 2021pfs is similar to that of SN 2011fe ($\Delta m_{15}(B)=1.12 \pm 0.05$ mag, \cite{Vinko2012A&A...546A..12V}). 

\begin{table}[htbp]
\centering
\caption{Photometric parameters of SN 2021pfs}
\begin{tabular}{ccccc}
\toprule
Band & $t_{\rm max}$ (MJD) & $m_{\rm peak} (mag)$ & Rise time (days) \\
\hline
\textit{U} & $59389.46\pm0.64$    &     $13.63\pm0.06$  &     $16.23\pm0.64$ \\ 
\textit{B} & $59391.49\pm0.34$    &     $13.82\pm0.03$  &     $18.27\pm0.35$ \\ 
\textit{g} & $59391.79\pm0.28$    &     $13.75\pm0.02$  &     $18.57\pm0.27$ \\ 
\textit{V} & $59391.87\pm0.24$    &     $13.77\pm0.02$  &     $18.65\pm0.25$ \\
\textit{r} & $59392.02\pm0.47$    &     $13.83\pm0.01$  &     $18.80\pm0.47$ \\ 
\textit{i} & $59391.85\pm0.09$    &     $14.28\pm0.01$  &     $18.63\pm0.11$ \\ 
\hline
\end{tabular}
\label{tab:phot_par}
\end{table}

\begin{figure}[ht!]
    \centering
	\includegraphics[width=6.in]{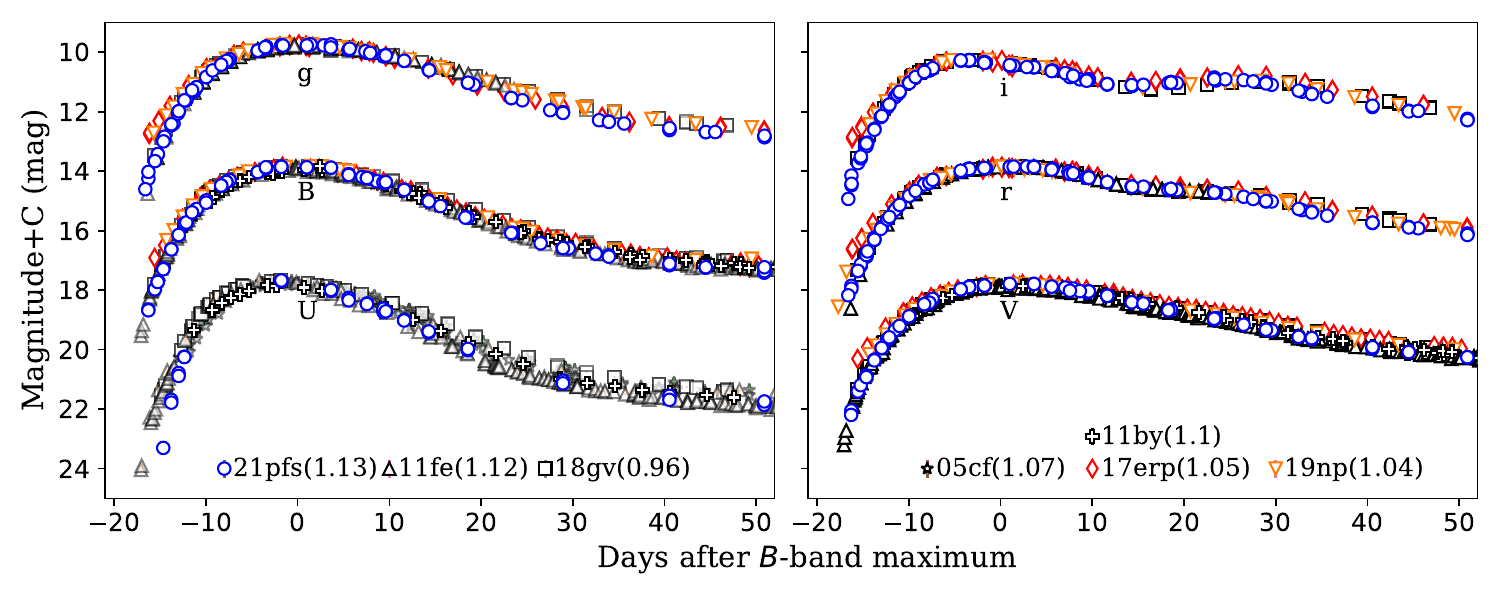}
	\caption{The near-peak light curves of SN 2021pfs are compared with those of several well-observed SNe Ia with similar $\Delta m_{15}(B)$ values, including SNe~2011fe, 2018gv, 2005cf, 2017erp, and 2019np. The light curves in different bands are vertically shifted to match their magnitudes at the \textit{B}-band maximum. The numbers in parentheses denote the $\Delta m_{15}(B)$ values of each SN Ia.
}
	\label{fig:col_lc}
\end{figure}

\begin{figure}[ht!]
    \centering
	\includegraphics[width=4.5in]{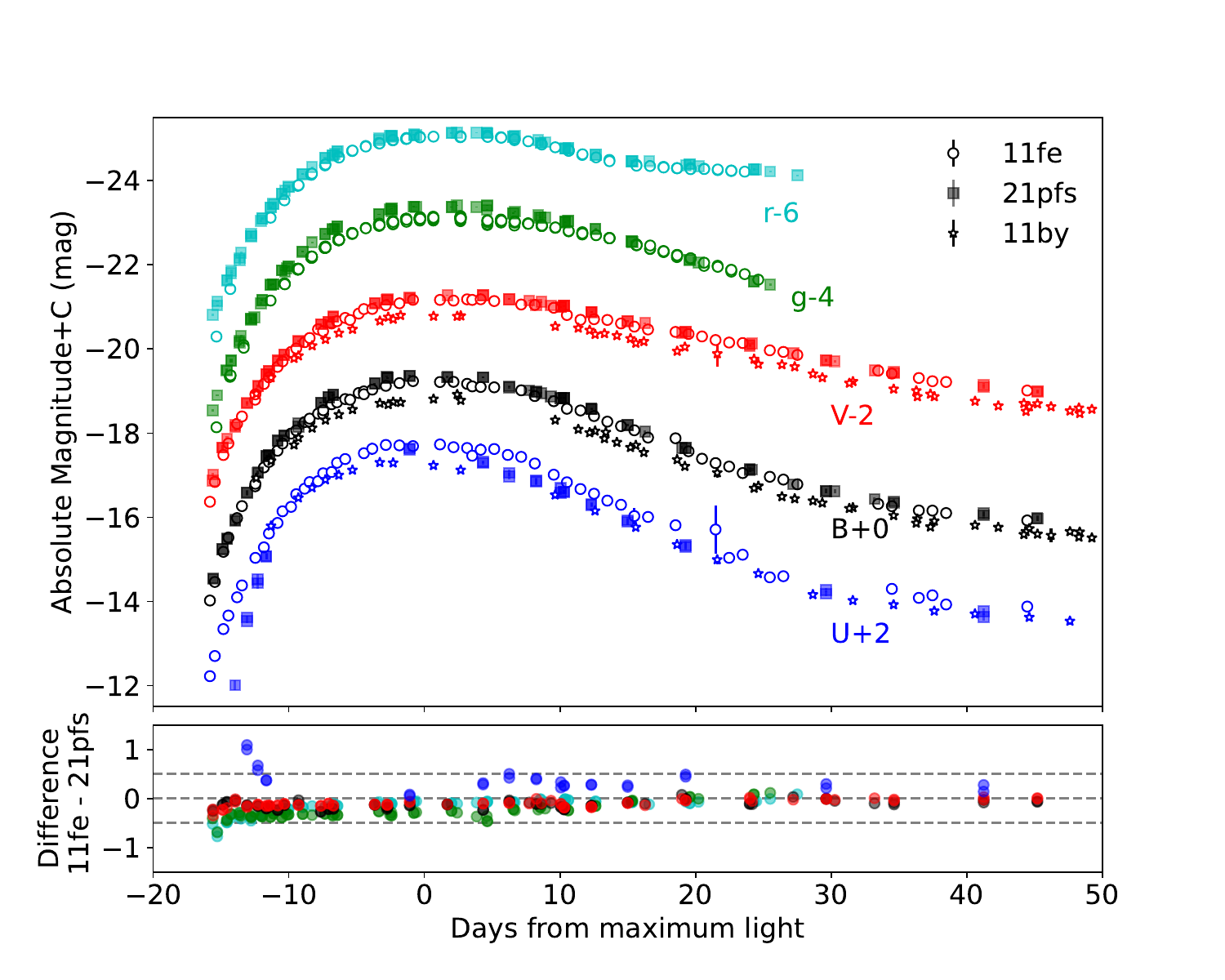}
	\caption{Absolute-magnitude \textit{UBVgr}-band light curves of SN 2021pfs (open circle) and SN 2011fe (filled square), together with the \textit{UBV}-band light curves of SN 2011by (open stars). The light curves in different bands are vertically shifted for clarity. The bottom panel plots the magnitude difference (SN 2011fe $-$ SN 2021pfs) in each band.
}
	\label{fig:abmag}
\end{figure}

In Figure \ref{fig:col_lc}, we compare the light curves of SN 2021pfs to those of several well-observed SNe Ia. For SNe without an early flux excess, we include SN 2011fe~\cite[$\Delta m_{15}(B)$=1.12;][]{Vinko2012A&A...546A..12V,Richmond2012JAVSO..40..872R,Munari2013fe,Zhang2016ApJ_11fe}, 
SN 2018gv~\citep[$\Delta m_{15}(B)$=0.96;][]{Yang2020ApJ},
SN 2005cf~\citep[$\Delta m_{15}(B)$=1.07;][]{Wang2009ApJcf},
and SN 2011by~\citep[$\Delta m_{15}(B)$=1.14;][]{Graham2015MNRAS.446.2073G}. For SNe with an early flux excess, we include SN 2019np~\citep[$\Delta m_{15}(B)$=1.04;][]{Sai2022MNRAS}, and SN 2017erp~\citep[$\Delta m_{15}(B)$=1.05;][]{Brown2019ApJerp}.
SN 2021pfs has even earlier coverage of its initial rise than most of these comparison objects. Overall, its photometric evolution closely resembles other normal SNe Ia presented for comparison, especially SN 2011fe. The early-time light curves of SN 2021pfs are more consistent with observational samples that show no early flux excess. Notably, at the very early stage, 
the $U-$band light curve of SN\,2021pfs appears to be remarkably fainter than that of SN\,2011fe (see Figure \ref{fig:abmag}).
This can also be interpreted as the rise time being faster at the shorter wavelengths and slower at longer wavelengths. A quantitative analysis of the rise times in different bands is given in Section \ref{sec:flt}.

\begin{figure}[ht!]
    \centering
	\includegraphics[width=\linewidth]{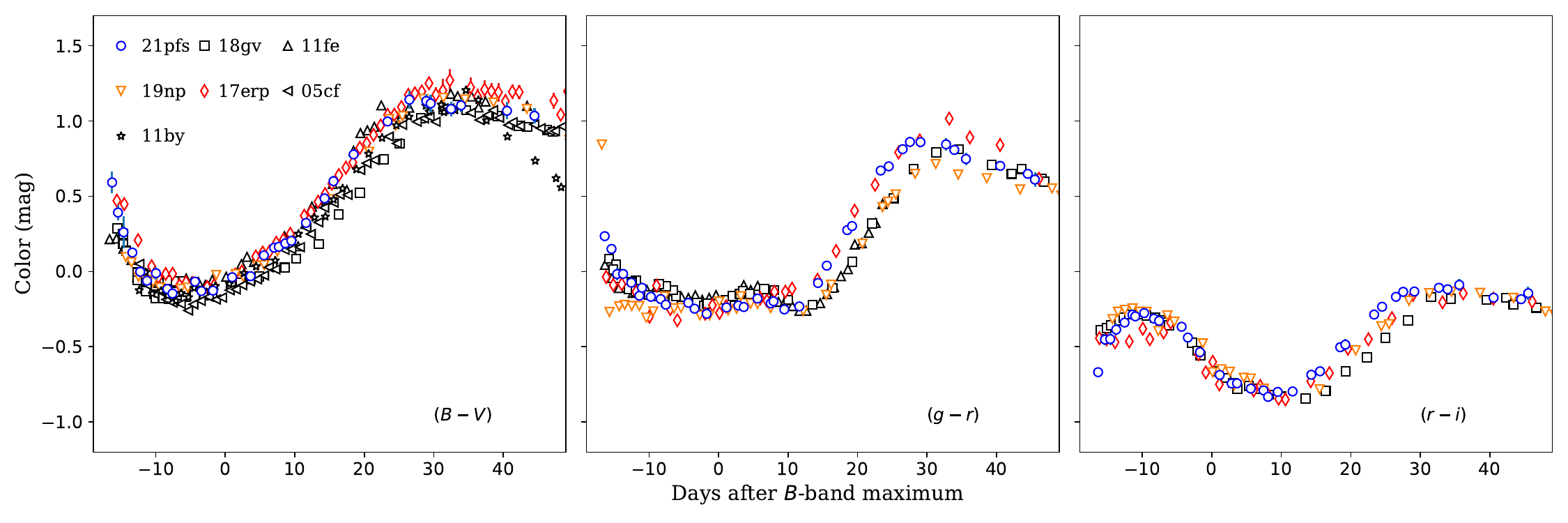}
	\caption{The color curves of SN 2021pfs are compared with other well-observed SNe Ia with similar $\Delta m_{15}(B)$, using the same comparison sample as in Figure \ref{fig:col_lc}.}
	\label{fig:color}
\end{figure}

Using the Galactic dust maps of~\cite{Schlegel1998ApJdustmag}  as recalibrated by~\cite{Schlafly2011ApJ}, we estimate a line-of-sight Milky Way extinction as $\rm A_{{V}}$ $\sim 0.077$ mag for SN 2021pfs. Adopting $\rm R_{V} = 3.1$, this corresponds to a Galactic color excess $E(B~-~V )_{\rm gal} \sim 0.025$ mag. The SNooPy2 fit indicates a host-galaxy reddening of $E(B - V )_{\rm host}$ = \ebvhost mag. Alternatively, using the empirical relation from~\cite{Phillips1999AJ} yields $E(B - V )_{\rm host}$ = $0.092~\pm$ 0.034 mag, consistent with the SNoopy2 result. Given an observed $B_{\rm max} - V_{\rm max}$ color of 0.05 $\pm$ 0.03 mag and an intrinsic color ${B-V}$ = $-0.067~\pm$ 0.012 mag. Here, the intrinsic color was obtained using Equation 7 from~\cite{Phillips1999AJ}. In this work, we adopt the SNoopy2 result. Thus, the total color excess is $E(B~-~V )_{\rm total}$ = 0.142 $\pm$ 0.060 mag. 

Figure \ref{fig:color} presents the color curves of SN 2021pfs compared to other SNe Ia with well-measured color curves. All colors are corrected for both Galactic and host-galaxy extinction. The color curves of SN 2021pfs are very similar to those of the comparison SNe Ia. At early times ($< -5$ days), its colors evolve more rapidly, the ${B-V}$ and ${g-r}$ colors are redder than those of others, whereas the ${r-i}$ color is bluer. Preivous studies found that most ${B-V}$ SNe Ia blue $B\!-\!V$ colors at early times belong to the SS spectral type of Branch classification, while early red events are exclusively classified as Branch CN or CL types \citep{Stritzinger2018ApJ}. The relatively red early colors of SN 2021pfs suggest it does not belong to the Branch SS subtype; as we discuss in Section \ref{sec:sepc}, its spectral features indicate a CN classification.

Adopting the SNooPy2 distance modulus ($\mu = 32.65 \pm 0.09$ mag) for SN 2021pfs, we can compare it to other distance estimates for the host galaxy.  The Tully-Fisher distance modulus of NGC 5427  is reported as $32.91 \pm 0.80$ mag \citep{Tully1988cng}, and the ``Sosies" method gives $32.34 \pm 0.83$ mag \citep{Terry2002A&A}. For subsequent analysis, we use an average $\mu~=$ \DM mag to construct the bolometric light curve. 
With this distance and the above extinction, the peak absolute magnitude of SN 2021pfs in \textit{B} band is estimated as \ABmag mag. This is very similar to the canonical absolute brightness of normal SNe Ia, and in particular to SN 2011fe \cite[$-19.21\pm0.15$ mag;][]{Richmond2012JAVSO..40..872R}. For SNe Ia with $\Delta m_{15}(B)$ $\sim$1.1 mag, the peak absolute magnitude is expected to be $\sim -19.3$ mag \citep{Phillips1999AJ,Wang2009HVNV}, consistent with our measurements. 

To further illustrate the similarity between SN 2021pfs and SN 2011fe, Figure \ref{fig:abmag} compares their absolute \textit{UBVgr}-band light curves, with the \textit{UBV}-band light curves of SN 2011by included for reference. After about $-$14 days, the differences in all bands remain within $\pm0.5$ mag, reaching their smallest values near maximum light. This demonstrates that the two SNe Ia have remarkably similar multi-band photometric evolution, particularly in the \textit{B} and \textit{V} bands. 

\begin{figure}[ht!]
    \centering
	\includegraphics[width=3.0in]{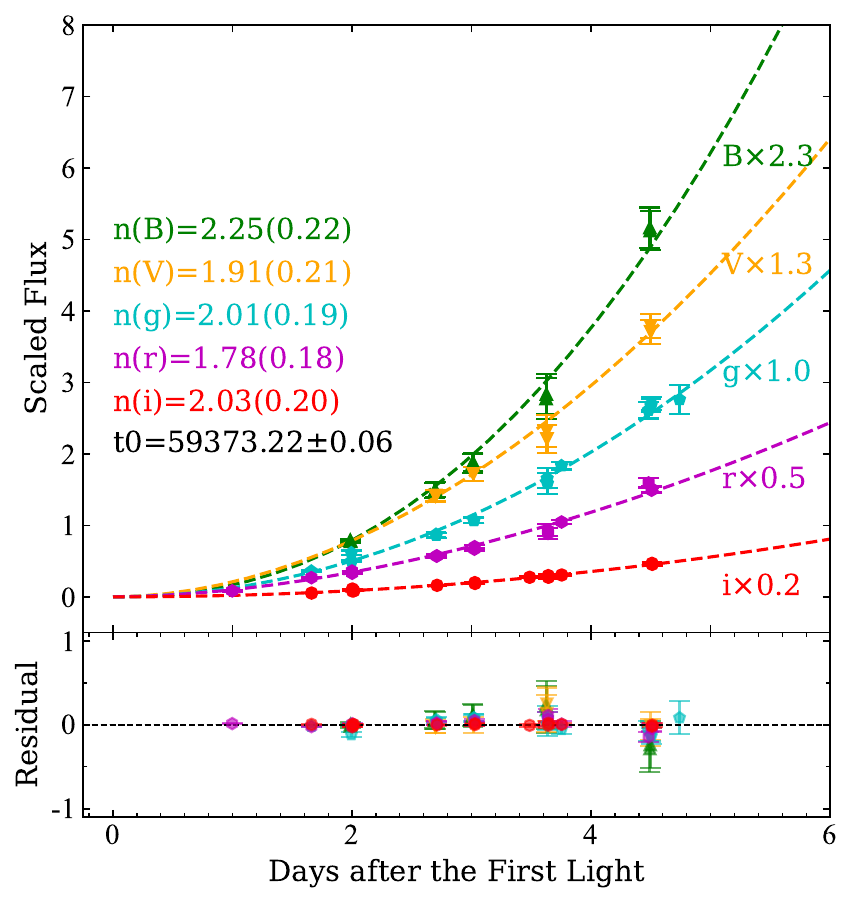}
	\caption{Fits to the early-time light curves of SN 2021pfs in the \textit{BVgri} bands. The dotted lines show the best-fit results obtained using the expanding fireball model (i,e., $f\propto(t-\rm FLT)^n$) \citep{Riess1999AJ}. The fitting residuals are shown in the bottom panel. For clarity, the light curves in different bands have been multiplied by constant factors.}
	\label{fig:FLT}
\end{figure}

\subsection{The First Light Time}\label{sec:flt} 
Early, high-cadence observations are necessary to precisely determine the First Light Time (FLT). Our follow-up observations of SN 2021pfs started at $\sim$ 0.1 day after the first alert. The last non-detection occurred at MJD 59372.3104, with a limiting magnitude of $\sim 16.87$ mag in the ZTF $r-$band \citep{Munoz-Arancibia2021TNSTR}. It is now well established that SNe Ia can exhibit a range of early rise behavior (power-law indices).
We fit the early-time multi-band data of this SN using a simple expanding fireball model, $f(t) \propto (t-\rm FLT)^\textit{n}$ model \citep{Riess1999AJ,iPTF16abc2018,Liu2024ApJ...973..117L}. Using data up to approximately $< -13$ days relative to the \textit{B}-band peak and assuming all bands share the same FLT, we obtain the best-fit indices of $n= 2.25\pm0.22$ in \textit{B}, $2.01\pm0.19$ in \textit{g}, $1.91\pm0.21$ in \textit{V}, $1.78\pm0.18$ in \textit{r}, $2.03\pm0.20$ in \textit{i} (Figure \ref{fig:FLT}). The common FLT from this fit is MJD $59373.22 \pm 0.06$ days, implying a \textit{B}-band rise time of \tbrise days, consistent with the average value for normal SNe Ia \citep{Zheng2017ApJ}. Given that our observations started at $\sim 1$ day after the explosion, the multi-band photometry of SN 2021pfs closely follows the $L \propto t^2$ law (as expected from an expanding fireball) within the first 5 days. This indicates no early flux excess in its optical light curves.  

\section{Optical Spectra}
\label{sec:sepc}
\begin{figure}
    \centering
	\includegraphics[width=3.in,height=4.in]{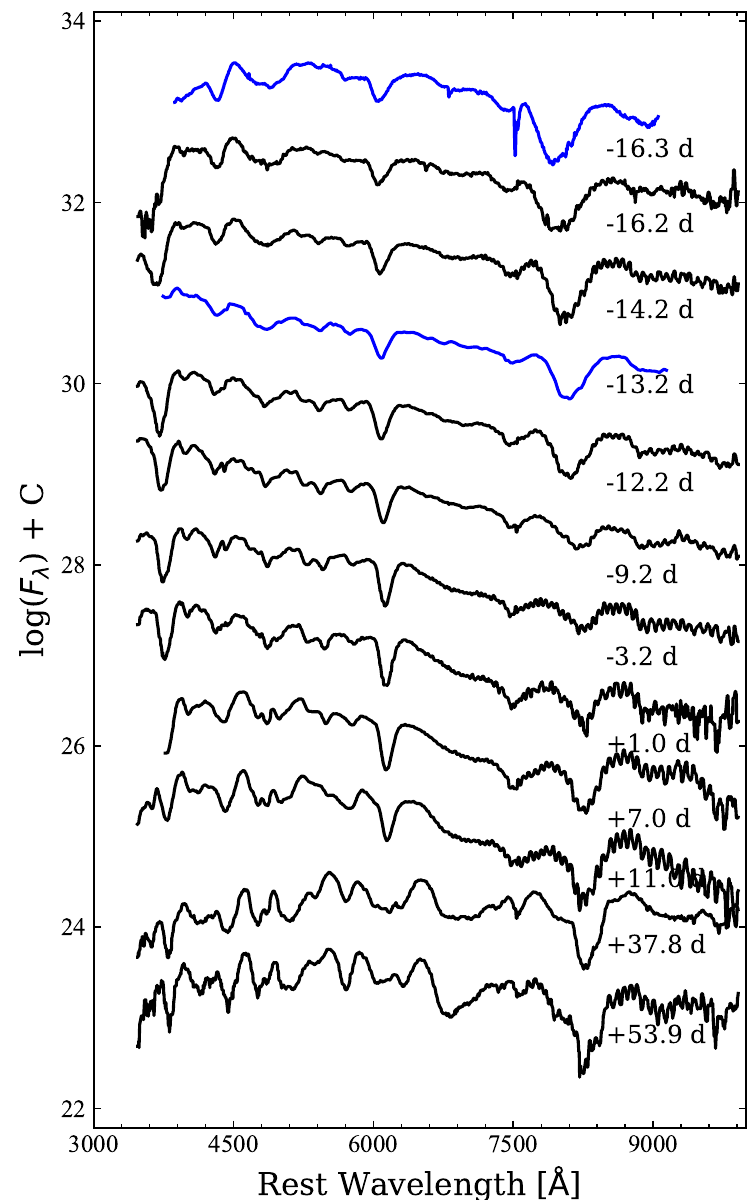}
	\caption{Optical spectra of SN 2021pfs obtained with LCO (black), and  archival spectra obtained from TNS (blue).
}
	\label{fig:spec}
\end{figure}

Figure \ref{fig:spec} displays the optical spectral sequence of SN 2021pfs from $-16.3$ days to $+53.9$ days relative to the $B$-band maximum. The spectra cover the optical range from ~3500 \AA~to ~10,000 \AA. The O {\sc i} $\lambda$7774 absorption feature is present in almost all spectra. Between $-12.2$ to $-3.2$ days, a weak C {\sc ii} $\lambda$6580 absorption line is also detectable. Before $\sim -12$ days, the Si {\sc ii} $\lambda$6355 absorption line shows an asymmetric profile, suggesting a possible high-velocity component on its blue wing. Additionally, the earliest spectra taken before –14 days shows a broad absorption trough between 4500 \AA~and 5000 \AA, which is a blend of multiple iron-group element lines. Below we discuss the spectral evolution in more detail.

\begin{figure}
    \centering
	\includegraphics[width=5.in]{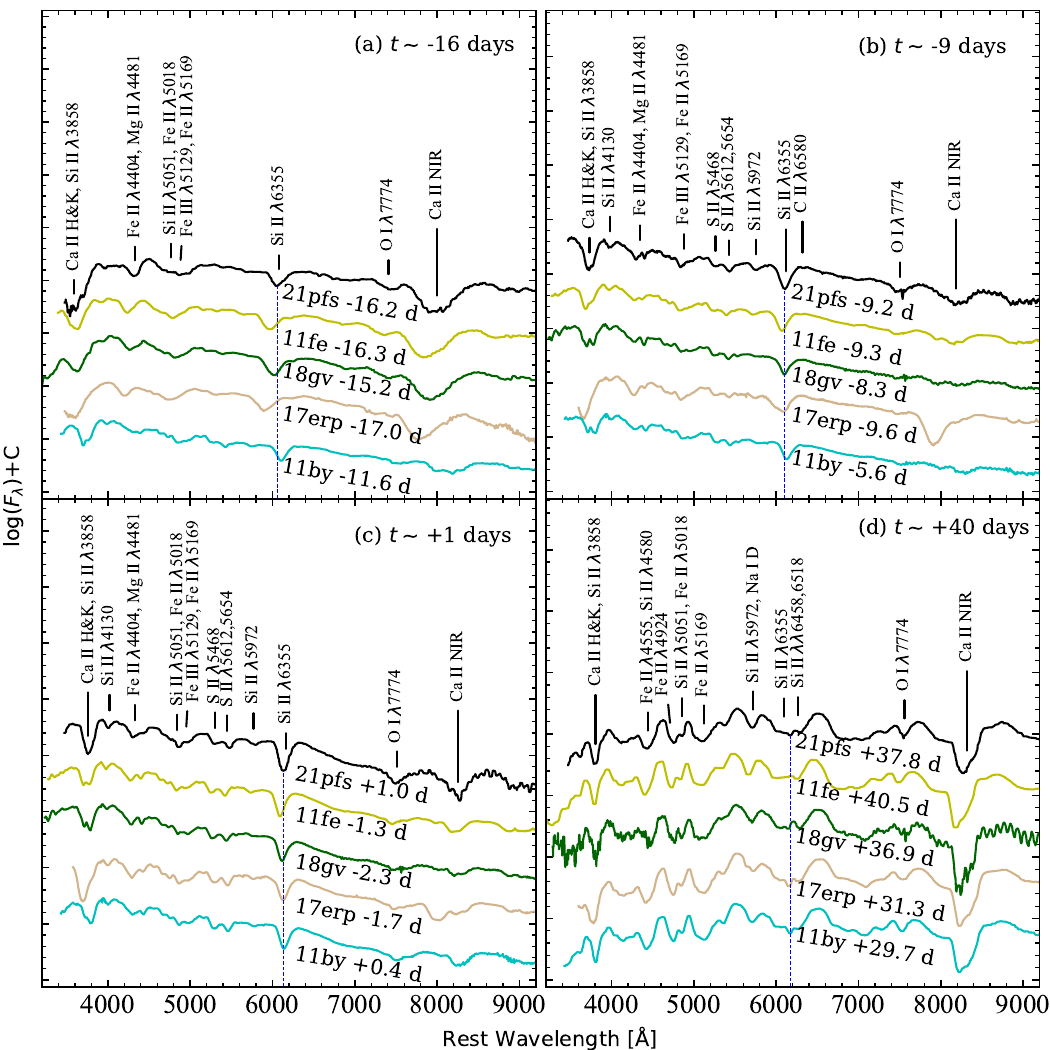}
	\caption{Spectral comparison of SN 2021pfs (black) at $-16.2, -9.2, +1.0, +37.8$ days with other SNe Ia at similar phases. The spectra of SN 2011fe~\citep{Zhang2016ApJ_11fe}, SN 2018gv~\citep{Yang2020ApJ}, SN 2017erp~\citep{Brown2019ApJerp}, SN 2011by~\citep{Graham2015MNRAS.446.2073G} are overplotted for comparison. Several absorption features (Ca {\sc ii} H\&K, Si {\sc ii}, Mg {\sc ii}, Fe {\sc ii}/{\sc iii}, O {\sc i}) are labeled. These spectra are shifted vertically for clarity. The blue vertical dotted line marks the absorption minimum of Si {\sc ii} $\lambda$ 6355 of SN 2021pfs in each panel.
}
	\label{fig:spec_comp}
\end{figure}
\subsection{Temporal Evolution of the Spectra}

Figure \ref{fig:spec_comp} compares the spectra of SN 2021pfs to those of other well-observed SNe Ia at four similar epochs. The comparison sample includes SN 2011fe~\citep{Zhang2016ApJ_11fe}, SN 2018gv~\citep{Yang2020ApJ,Burke2025ApJ...994...87B}, and SN 2011by~\citep{Graham2015MNRAS.446.2073G}, which show no early flux excess, as well as SN 2017erp \citep{Brown2019ApJerp,Burke2025ApJ...994...87B}, which exhibits an early excess. The $-$16.2 day spectrum of SN 2021pfs is dominated by P-Cygni absorption profiles from Si {\sc ii}, Fe {\sc ii}, Mg {\sc ii}, Fe {\sc iii}, O {\sc i}, and Ca {\sc ii}. Its spectral continuum and line profiles closely resemble those of SN 2011fe at $-$16.3 days and SN 2018gv at $-$15.2 days, except that the blended feature around 4500 \AA~differs from those two SNe (see Figure \ref{fig:spec_comp}(a)). Furthermore, the unburned O {\sc i} $\lambda$7774 absorption lines in SN 2021pfs are slightly stronger than in SN 2011fe/SN 2018gv. The absorption minimum of the Si {\sc ii} $\lambda$6355 line in SN 2021pfs is redshifted relative to those in SN 2011fe and SN 2018gv, indicating that the photospheric expansion velocities of SN~2021pfs at early times are lower.

By $\sim -$9 d (Figure \ref{fig:spec_comp}(b)), as the photosphere recedes and expansion velocity declines, we see more distinct spectral features. The blended absorption line of Fe {\sc ii, iii} and Si {\sc ii} near 4500 \AA~has now partially resolved into individual lines. Additional absorption lines of intermediate-mass elements (IMEs) become apparent, including the characteristic W-shaped S {\sc ii} doublet and Si {\sc ii} $\lambda$5972 line. Meanwhile, Si {\sc ii} $\lambda$4130 and $\lambda$6355 absorption lines have strengthened with time, whereas the Ca {\sc ii} NIR triplet absorption has weakened, though it remains stronger in SN 2021pfs than in SN 2011fe and SN 2018gv. A subtle C {\sc ii} $\lambda$6580 absorption line is  detected in SN 2021pfs at  t$\sim$$-$9.2 days, indicating some unburned carbon is still present in the outer ejecta. The O {\sc i} $\lambda$7774 absorption lines also appear slightly stronger than at $-$16 days. Overall, the spectrum of SN 2021pfs is remarkably similar to that of SN 2011fe at t$\sim$$-$9 days. The main differences are that SN 2021pfs exhibits stronger Ca {\sc ii} H\&K and NIR lines than SN 2011fe.

Around the maximum light (Figure \ref{fig:spec_comp}(c)), the spectra show the typical SN Ia features. The Si {\sc ii} $\lambda$6355 absorption line remains very strong, while the Si {\sc ii} $\lambda$5972 absorption line has begun to weaken slightly. The profile of the Ca {\sc ii} HK line is similar to what it was $\sim$8 days earlier, and it is still somewhat deeper than in SN 2011fe or SN 2018gv, more closely resembling the spectrum of SN 2017erp. The Ca {\sc ii} NIR triplet absorption lines also retains a different profile in SN 2021pfs compared to the others. The other IME lines (Si II, S II, etc.) in SN 2021pfs are quite similar to those in the comparison SNe Ia at maximum. At around the maximum light, the main differences between SN 2021pfs and SN 2011fe are still the stronger Ca {\sc ii} H\&K and NIR absorption features in SN 2021pfs.

In the +37.8 days spectrum (Figure \ref{fig:spec_comp}(d)), the O {\sc i} $\lambda$7774 line is still visible, which is also the case in the +40.5 days spectrum of SN 2011fe. The Ca {\sc ii} NIR absorption feature, which has faded around the peak, has strengthened again and now dominates the late-time spectrum. At this epoch, the spectrum of SN 2021pfs closely resembles those of the comparison SNe Ia.

In summary, the spectral evolution of SN 2021pfs is very similar to that of SN 2011fe and SN 2018gv, especially at the early epoch. The main differences are confined to the Ca II H\&K and NIR lines, and the broad absorption feature of iron-group elements at early times. We also detect unburned carbon (C {\sc ii} $\lambda$6580) in SN 2021pfs from $\sim -12$ until $\sim -3$ days. 

As the photosphere recedes, the Si {\sc ii} $\lambda$6355 absorption line becomes deeper and narrower, while 
Si {\sc ii} $\lambda$5972 lines are detectable in spectra. We measured the pseudo-equivalent widths (pEWs) of Si {\sc ii} $\lambda$5972 and Si {\sc ii} $\lambda$6355 absorption lines from the pre-maximum spectra of SN 2021pfs; the results are listed in Table \ref{tab:pew}. In the +1.0 day spectrum, the pEW($\lambda$5972) is 19.26 $\pm$ 0.2 Å and pEW($\lambda$6355) is 117.03 $\pm$ 1.1 Å. According  to the classification scheme of~\cite{Branch2006PASP}, these values place SN 2021pfs in the CN subclass. Using the probabilistic method of \cite{Burrow2020ApJ...901..154B}, we estimate $\sim 89\%$ probability that SN 2021pfs is CN. We also compute the line-strength ratio $R$(Si {\sc ii})= pEW($\lambda$5972)/pEW($\lambda$6355), which is a proxy for the photospheric temperature~\citep{Nugent1995ApJ}, with higher values corresponds to lower photospheric temperatures in SNe Ia. For SN 2021pfs around maximum, we find $R$(Si II) = $0.16 \pm 0.02$. 
The $R$(Si {\sc ii}) of SN 2021pfs is comparable to that of SN 2018oh ($R$(Si {\sc ii})= $0.15 \pm 0.04$,~\citealt{Li20190h}) and lower than that of SN 2021fe ($R$(Si {\sc ii})= $0.23 \pm 0.02$,~\citealt{Graham2015MNRAS.446.2073G}). This suggests that the photospheric temperature of SN 2021pfs around maximum light is similar to that of SN 2018oh and higher than that of SN 2011fe. 


\begin{figure*}[!t]
    \centering
    \includegraphics[width=6in]{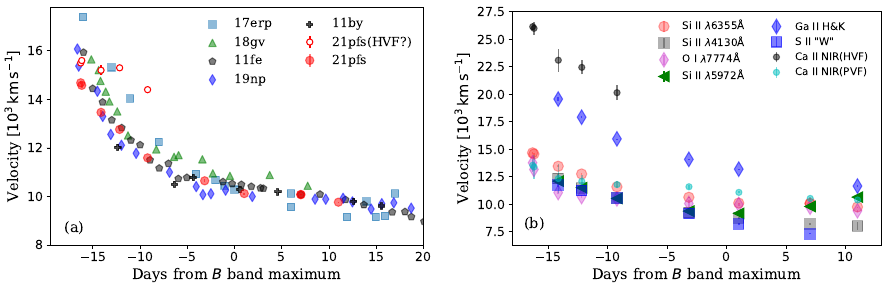}
    \caption{(a): Ejecta velocity of Si {\sc ii} $\lambda$6355 in SN 2021pfs compared to other SNe Ia: SN 2011fe \cite{Zhang2016ApJ_11fe}, SN 2018gv \cite{Yang2020ApJ}, SN 2017erp \cite{Brown2019ApJerp}, SN 2019np \cite{Sai2022MNRAS}, and SN 2011by~\cite{Graham2015MNRAS.446.2073G}. (b): Velocity evolution of various absorption lines in SN 2021pfs: Si {\sc ii} $\lambda$4130, 5972, 6355, O {\sc i} $\lambda$7774, Ca {\sc ii} H\&K, NIR triplet (showing both photospheric component and HVF), and S {\sc ii}.}
    \label{fig:vc}
\end{figure*}

\startlongtable
\begin{deluxetable*}{lCCC}
\tablecolumns{4}
\tablewidth{0pc}
\tabletypesize{\scriptsize}
\tablecaption{Pseudo-equivalent width of \ion{Si}{2} absorption lines in SN 2021pfs.}
\tablehead{
\colhead{MJD}
&\colhead{Phase\tablenotemark{a} (days)}
&\multicolumn{2}{c}{Pseudo-equivalent width ($\mathrm{\AA}$)}\\
\cline{3-4}
\colhead{}
&\colhead{}
&\colhead{\ion{Si}{2} $\lambda5972$}
&\colhead{\ion{Si}{2} $\lambda6355$}
}
\startdata
59375.20 & -16.3 & \nodata     & 98.1\pm3.4  \\
59375.31 & -16.2 & \nodata     & 81.3\pm6.3  \\
59377.35 & -14.2 & 18.1\pm1.6& 103.6\pm2.9 \\
59379.32 & -12.2 & 25.4\pm0.6& 108.4\pm1.0 \\
59382.28 & -9.2  & 23.9\pm0.7& 103.8\pm2.6 \\
59388.32 & -3.2  & 24.3\pm1.1& 114.1\pm0.4 \\
59392.51 & +1.0  & 19.3\pm0.2& 117.0\pm1.1 \\
\enddata
\tablenotetext{a}{Relative to the $B$-band peak maximum (MJD 59391.49).}
\label{tab:pew}
\end{deluxetable*}

\subsection{The Photospheric Expansion Velocity}
After correcting for host galaxy redshift to all spectra, we measured the Doppler velocities of several important absorption lines using Gaussian profile fits (with Monte Carlo uncertainty estimation \citep{Wang2009ApJcf,Zhao2015ApJS,Zhao2016ApJ}. For Si {\sc ii} $\lambda$6355, $\lambda$4130, $\lambda$5972, S {\sc ii}, O {\sc i} $\lambda$7774, Ca {\sc ii} H\&K lines, a single Gaussian provided a good fit to the line minimum. For the Ca {\sc ii} NIR triplet at early times (which often shows a high-velocity feature, HVF) and the Si {\sc ii} $\lambda$6355 line, multiple Gaussians were used to decompose the components. We compared the velocity of Si {\sc ii} $\lambda$6355 line in SN 2021pfs with other SN Ia samples (Figure \ref{fig:vc} (a)). At $-16.3$ days, the photospheric velocity measured from the Si {\sc ii} $\lambda$6355 was $ 14640 \pm 130 \rm\, km\, s^{-1}$, which is comparable to the velocity of O {\sc i} $\lambda$7774 line ($ 14000 \pm 250 \rm\, km\, s^{-1}$) at that time. This early Si {\sc ii} $\lambda$6355 velocity is lower than that of other comparison SNe Ia. Around the \textit{B} band maximum light, the Si {\sc ii} $\lambda$6355  velocity in SN 2021pfs had decreased to 10120 $\pm$ 140 km s$^{-1}$, the same as that of SN 2011fe at peak. According to the classification criteria of \cite{Wang2009HVNV}, we categorize SN 2021pfs as an NV SNe Ia. We further calculate the Si {\sc ii} velocity gradient between $+$1.0 and $+$11.0 days, finding $\dot v_{\rm Si\,II} = 26.5 \pm 5.0$ km s$^{-1} \rm day^{-1}$. This places SN 2021pfs in the LVG subclass according to the \cite{Benetti2005ApJ} scheme. For reference, SN 2011fe is also an NV and LVG SN Ia \citep{Parrent2012ApJ...752L..26P,Zhang2016ApJ_11fe}.

Observations suggest that $\sim 95\%$ of SNe Ia show a HVF Ca {\sc ii} NIR component in spectra earlier than $-5$ days \citep{Maguire2014MNRAS}. SN 2021pfs conforms to this trend. As seen in Figure \ref{fig:vc}(b), the Ca {\sc ii} velocities are significantly higher than those of Si {\sc ii} and other IME elements. The Ca {\sc ii} NIR HVF in SN 2021pfs exceeds  $\sim 11500 \rm\, km\, s^{-1}$ above the Si {\sc ii} photospheric velocity, indicating that the material forming the Ca {\sc ii} HVF is likely the outermost layer of the ejecta. We notice that the Si {\sc ii} $\lambda$6355 line profiles in the earliest spectra (before $-$12 days) are asymmetric, suggesting a potential Si {\sc ii} HVF as well. Using a multi-Gaussian fit, we find that in the $-$15.5 day spectrum of SN 2021pfs, a high-velocity Si {\sc ii} component is centered at $ 15400 \pm 100 \rm\, km\, s^{-1}$, while the low-velocity component is at $ 11300\pm 100 \rm\, km\, s^{-1}$. The difference confirms the presence of a Si {\sc ii} HVF in the early spectra of SN 2021pfs. 


\section{Discussions}\label{sec:discuss}
\subsection{Nickel Mass-Luminosity Discrepancy and Ejecta Properties}\label{sec:Ni}
The photometric observations of SN 2021pfs (covering \textit{uUBgVrizy}, roughly 300-1100 nm) allow us to construct a detailed quasi-bolometric light curve. Using SNooPy2, we computedthe spectral energy distribution and quasi-bolometric light curve of SN 2021pfs, following the same procedure used for SN 2018oh \citep{Li20190h}. We assumed that the ultraviolet and near‑infrared contributions are each about 5$\%$ of the bolometric flux \citep{Wang2009ApJcf,Li20190h,Zhang2016ApJ_11fe}. 
Under these assumptions, we find a peak bolometric luminosity of  \Lbolo $\rm erg~s^{-1}$, occuring at MJD 59390.96 ($\sim$ 0.53 days before the $B$-band maximum). For comparison, SN 2011fe reached a peak of $\sim 1.13 \times 10^{43} \rm erg~s^{-1}$, higher than SN 2021pfs \citep{Zhang2016ApJ_11fe}).

\begin{figure}
    \centering
	\includegraphics[width=3in]{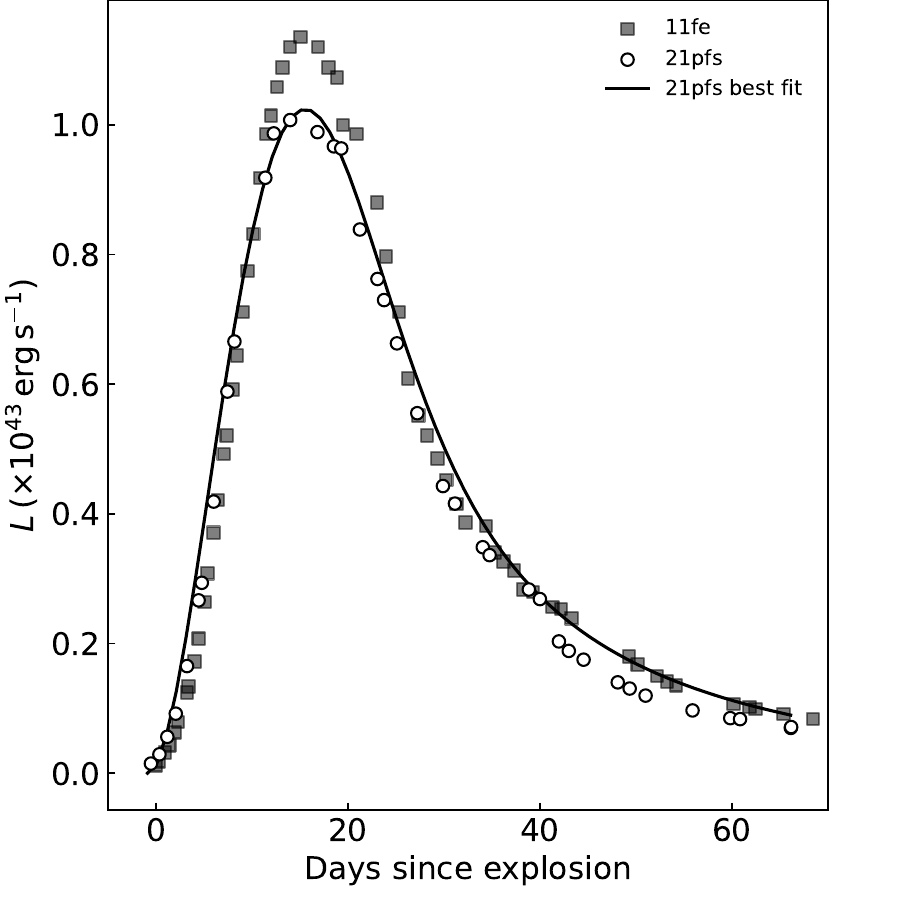}
	\caption{Quasi-bolometric light curve of SN 2021pfs was created utilizing the SNooPy2 \citep[open circle]{Burns2011AJ}. The solid black line is the best-fit result obtained using the Arnett model \citep{Arnett1982}. The square represents the quasi-bolometric light curve of SN 2011fe.}
	\label{fig:bolo}
\end{figure}

By fitting the Arnett model \citep{Arnett1982,Chatzopoulos2012,Li20190h} to the bolometric light curve, we estimate the mass of radioactive $^{56}$Ni  synthesized during the explosion, along with other ejecta parameters. Producing more $^{56}$Ni yields a brighter peak and a broader light curve because the increased opacity prolongs diffusion time. We employed the semi-analytic \texttt{minim} \citep{Chatzopoulos2013ApJ} code to fit the quasi-bolometric light curve with a constant opacity Arnett model. 
The model parameters we solved for were: (1) FLT $t_0$; (2) the $^{56}$Ni mass formed during the explosion $M_{\rm Ni}$; (3) the light-curve diffusion timescale $t_{\rm lc}$, and (4) the timescale of gamma-ray leakage $t_{\gamma}$ \citep{Chatzopoulos2012,Chatzopoulos2013ApJ}. The best-fit result (Figure \ref{fig:bolo}) gives $t_0 \approx -0.53$ days (relative to our earliest observation), $M_{\rm Ni} = 0.48 \pm 0.05~M_{\odot}$, $t_{\rm lc}= 14.47 \pm 0.02$ days, and $t_{\gamma} = 35.41 \pm 1.73$ days. This $M_{\rm Ni}$ of SN 2021pfs is lower than the $M_{\rm Ni}$ for SN 2018oh \citep[$\sim 0.55~M_{\odot}$;][]{Li20190h} and SN 2011fe \citep[$\sim 0.57~M_{\odot}$;][]{Zhang2016ApJ_11fe}. Using the method of \cite{Li20190h}, we deduce an effective optical opacity of $0.21 \pm 0.05 \rm~cm^{2}~g^{-1}$. 
Combining the above parameters with Equation 1 of \cite{Li20190h}, we estimate an ejecta mass $M_{\rm ej}~=~0.94 \pm 0.09~M_{\odot}$ and a kinetic energy $E_{\rm k}~=~(8.04 \pm 0.18) \times 10^{50} \rm~erg$ for SN 2021pfs. These values suggest SN 2021pfs was a typical SN Ia in terms of ejecta mass and energy, though perhaps on the lower side of the Ni mass distribution.

\begin{table}
	\centering
	\setlength{\tabcolsep}{8mm}
	\caption{Main parameters of SN 2021pfs}
	\begin{tabular}{cccccll}
		\hline
        \hline
		Parameter                                       &SN 2021pfs               &Unit  \\
		\hline     
		$\Delta m_{15}(B)$                              &\dm15                    & mag  \\
		$t(B_{\rm max})$                                &\tbmax                   & days \\
		$M(B_{\rm max})$                                &\mbmax                   & mag  \\
        $t_0$                                           &\t0                      & days \\
  	    Rise time(B)                                    &\tbrise                  & days \\
        Distance Modulus                                &\DM                      & mag  \\
        $E(B-V)_{\rm MW}$                               & $\approx 0.025$         & mag  \\
        $E(B-V)_{\rm host}$                             & \ebvhost                & mag  \\
        Absolute Magnitude(B)                           &\ABmag                   & mag  \\
        $L_{\rm bol}^{\rm max}$                         & \Lbolo                  & $\rm erg\, s^{-1}$\\
        $M_{\rm ^{56}Ni}$                               & \nimass                 & $\rm M_\odot$\\
		$\upsilon_{\rm 0}$(Si {\sc ii})$_{\rm max}$     &10120 $\pm$ 140          & km s$^{-1}$ \\
        $\dot{\upsilon}$(Si {\sc ii})                   &26.50 $\pm$ 5.00         &km s$^{-1} \rm \, day^{-1}$ \\
		\hline
	\end{tabular}
\label{tab:mianpar}
\end{table}

The Arnett model depends on $t_{\rm lc}$ and the optical opacity, which has large uncertainties \citep{Arnett1982,Chatzopoulos2012}. Moreover, the degeneracy between optical opacity and ejecta mass may introduce systematic biases in mass estimation \citep{Bora2024PASP..136i4201B}. Given the limitations of the Arnett model, we attempted to estimate the ejecta mass and kinetic energy using the gamma-ray leakage timescale $t_{\gamma}$ combined with the gamma-ray opacity $\kappa_\gamma$ \citep{Bora2024PASP..136i4201B}. Here, $\kappa_\gamma$ was taken as 0.025 $\rm cm^{2}~g^{-1}$ from \cite{Guttman2024MNRAS.533..994G}. The derived ejecta mass is $M_{\rm ej}~=~0.81 \pm 0.08~M_{\odot}$, and the corresponding kinetic energy is $E_{\rm k}~=~(4.94 \pm 0.55) \times 10^{50} \rm~erg$. Both the ejecta mass and kinetic energy derived using Bora's method are smaller than those obtained from the Arnett model. The gamma-ray leakage method relies on fitting the late-time light curve tail, where the radiation is dominated by $^{56}$Co decay and is less affected by early mixing effects and energy injection processes, making its results relatively more robust \citep{Bora2024PASP..136i4201B}.

The interplay between ejecta mass, kinetic energy, and density profile determine the photon diffusion timescale and thus regulate the efficiency of the release of decay energy from $^{56}$Ni. 
The distribution of $^{56}$Ni in the ejecta also plays a crucial role: if a significant fraction of the Ni is mixed into the outer layers, it can enhance the early luminosity (by providing high-energy photons that escape earlier), while a more centrally concentrated Ni distribution delays the release of energy, making the peak later \citep{Ni56Magee2020}. Multi-dimensional simulations further indicate that asymmetries in the Ni distribution can lead to deviations from the expectations of spherically symmetric models \citep{Kasen2009Natur.460..869K,Maeda2010Natur.466...82M}. In particular, an asymmetric explosion can hide some Ni-rich material from direct view, affecting the inferred Ni mass from the light curve.

If radioactive nickel is present in the outer ejecta, the gamma-ray photons from Ni decay can escape more readily, contributing excess UV/blue flux at early times \citep{Ni56Magee2020}. However, the thermalization efficiency of outer ejecta layers is relatively low, so not all the decay energy is converted into optical radiation; consequently, the observed luminosity might be lower than expected from a fully trapped scenario \citep{DarkPhase2013,Ni56Magee2020}. Additionally, nickel-rich regions deep in the ejecta have high line opacity (due to a forest of Fe-group lines), which can significantly slow the diffusion of energy \citep{Kasen2010,Pinto2000ApJ...530..757P}. All these factors can cause the simple one-to-one relation  between nickel mass and peak luminosity (as per Arnett’s rule) to break down to some degree. 

Another consideration is interactions between the ejecta and dense circumstellar material (CSM) or shock cooling emission from a companion (such as in the two-component model of SN 2021agco, \citealp{SN2021agcoYan,Kasen2010}). If an SN Ia explodes inside a dense CSM or has a strong collision with the companion’s envelope, additional early light can be generated, potentially leading to an overestimate of nickel mass if one assumes the luminosity is purely from radioactive decay \citep{Miller2020SN2019yvq}. Furthermore, the simplifying assumption that the peak luminosity equals the instantaneous $^{56}$Ni decay power may  not hold exactly, especially if the timescales for energy deposition and diffusion are dismatched or if there is a steep radial density gradient in the ejecta \citep{Kasen2017}. 

In summary, while our Arnett-model analysis provides a reasonable estimate of SN 2021pfs’s Ni mass and ejecta properties, one should keep in mind the above caveats. The slightly lower Ni mass (and thus peak luminosity) of SN 2021pfs compared to SN 2011fe could be influenced by its higher progenitor metallicity (see Section \ref{sec:hostgalaxy}) and any differences in explosion asymmetry or Ni distribution.

\begin{figure}
	\begin{minipage}[t]{0.32\linewidth}  
		\centering 
		\includegraphics[width=2.5in]{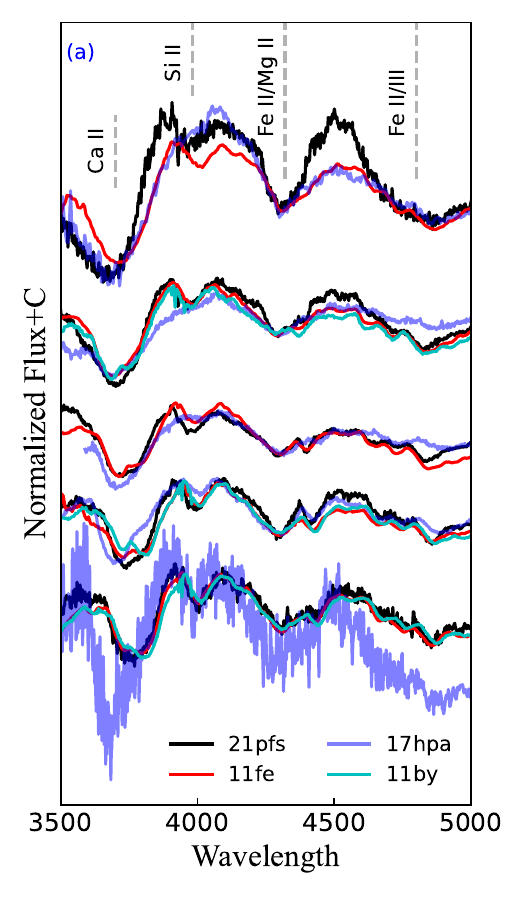}   
	\end{minipage} 
	\begin{minipage}[t]{0.32\linewidth}   
		\includegraphics[width=2.5in]{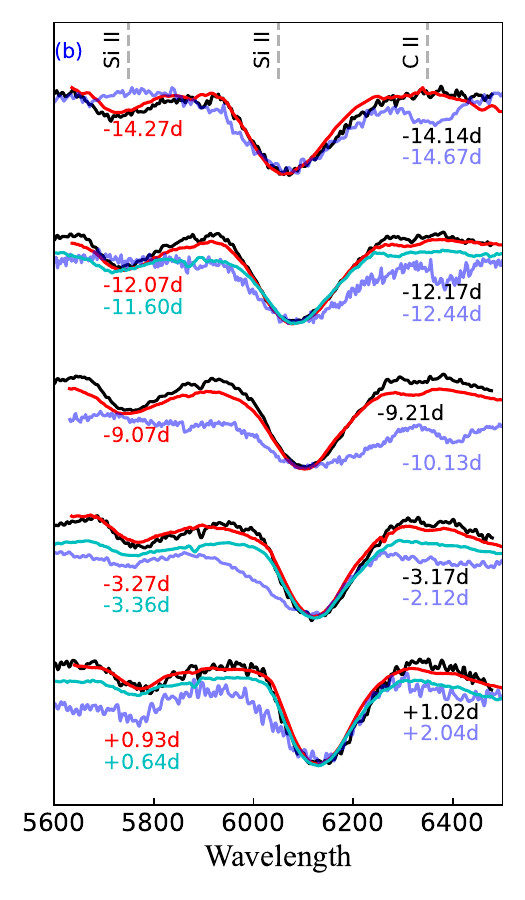}   
	\end{minipage} 
	\begin{minipage}[t]{0.32\linewidth}    
		\includegraphics[width=2.5in]{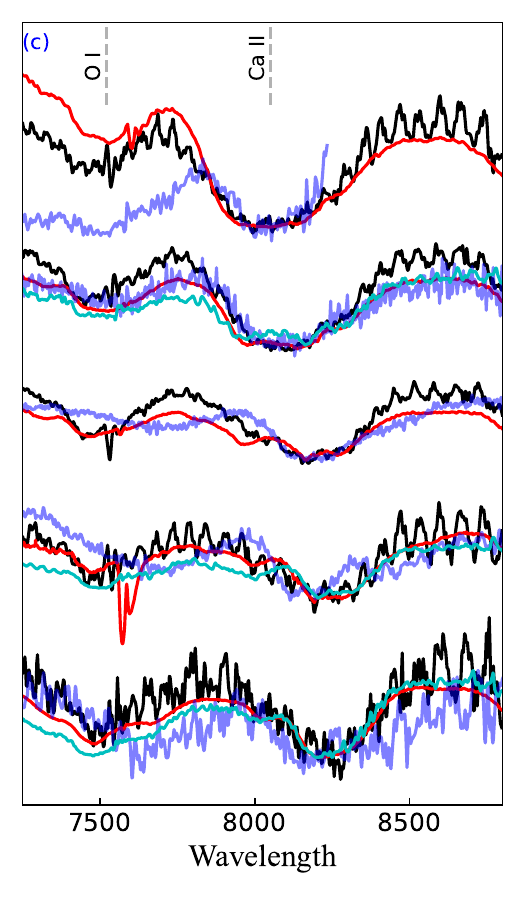}   
	\end{minipage} 
	\caption{Early-time spectral comparison of SN 2021pfs (black), with SN 2011fe (red; \citealt{Parrent2012ApJ...752L..26P}), SN 2017hpa (blue; \citealt{Zeng2021hpa}), and SN 2011by (cyan; \citealt{Graham2015MNRAS.446.2073G}). 
    All spectra are normalized at the absorption minimum of Fe {\sc ii} $\lambda$4404 (a), Si {\sc ii} $\lambda$6355 (b), and Ca {\sc ii} NIR (c). Panel~(a) focuses on the Fe~{\sc ii} $\lambda4404$ region, panel~(b) on Si~{\sc ii} $\lambda6355$ (with the C~{\sc ii} $\lambda6580$ feature indicated), and panel~(c) on the Ca~{\sc ii} NIR triplet. The phase of each spectrum is annotated in panel~(b).}
	\label{fig:C6580}
\end{figure}

\subsection{Are SN 2021pfs and SN 2011fe twins?}
The striking similarities between SN 2021pfs and SN 2011fe invite the question of whether these two could be considered ``twin” SNe Ia. As shown in Figure \ref{fig:abmag}, their absolute-magnitude light curves are extremely alike in most bands. Although the similarity in the \textit{U} and \textit{i} bands is less pronounced, the difference in peak remains within 0.5 mag. Both supernovae have nearly the same $\Delta m_{15}(B)$ and \textit{B}-band absolute magnitude, and both are  NV/LVG SNe Ia. 

\begin{figure}
    \centering
	\includegraphics[width=5in]{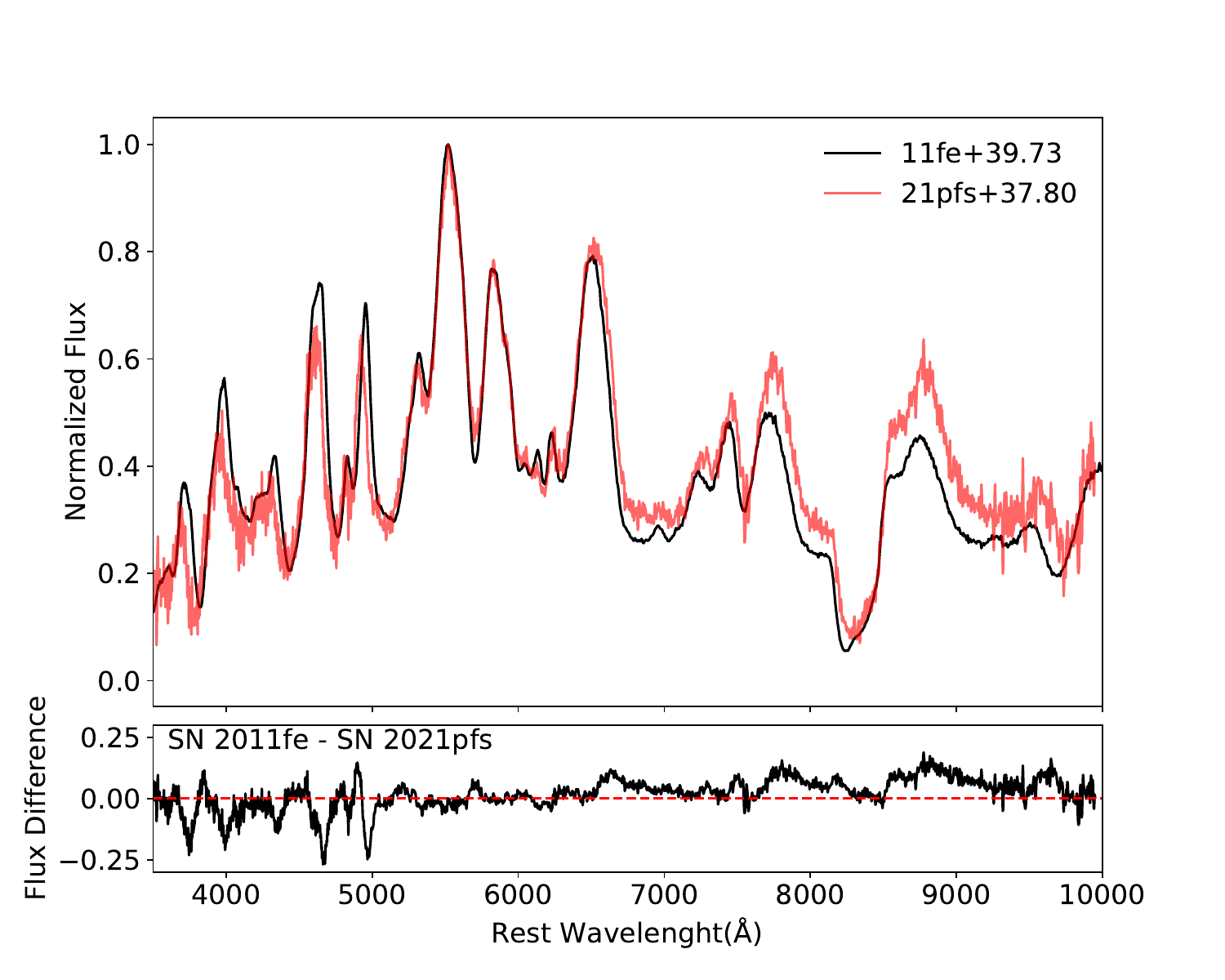}
	\caption{Top: Late-time spectra of SN 2021pfs (red) at +37.80 days and SN 2011fe (black) at +39.73 days after peak brightnesses. Bottom: Difference spectrum (SN 2011fe minus SN 2021pfs) after the spectra are normalized and aligned at their peak flux.}
	\label{fig:spec38}
\end{figure}

To further investigate this, we performed a detailed comparison between the pre‑maximum spectra of SN 2021pfs and those of SN 2011fe \citep{Zhang2016ApJ_11fe}, SN 2011by which has been studied as a “twin” to SN 2011fe \citep{Graham2015MNRAS.446.2073G,Foley2020MNRAS.491.5991F}, and the carbon‑rich SN Ia SN 2017hpa \citep{Zeng2021hpa} at similar epochs, focusing on the profiles of several prominent absorption features, including Si {\sc ii} $\lambda$4130, $\lambda$5972, $\lambda$6355, Ca {\sc ii} H\&K and the NIR triplet. The comparison is presented in Figure \ref{fig:C6580}. The spectra of SN 2021pfs closely resemble those of SN 2011fe, including the C{\sc ii} absorption feature near $\lambda$6580. Both SNe exhibit only weak C {\sc ii} $\lambda$6580 absorption feature, in stark contrast to the prominent feature observed in the carbon‑rich SN 2017hpa (Figure \ref{fig:C6580}(b)). Relative to SN 2011by, SN 2021pfs shows a closer match to SN 2011fe in the line profiles of the major absorption features. 
The late-time spectrum (+38 days) of SN 2021pfs is also remarkably similar to that of SN 2011fe at +40 days (Figure \ref{fig:spec38}, top panel), , with the difference spectrum showing only minor deviations (Figure 12, bottom panel) All these points support the idea that SN 2021pfs is a very close analog of SN 2011fe. 

However, “twins” are rarely absolutely identical. We do find some subtle differences between SN 2021pfs and SN 2011fe. As discussed earlier, SN 2021pfs’s early $U$-band light curve rises faster. SN 2021pfs also had a somewhat lower bolometric peak luminosity and Ni mass than SN 2011fe (Section \ref{sec:Ni}). In the spectra, SN 2021pfs had systematically lower Si {\sc ii} velocities  during the early stage (though by maximum light they converged) (see Figure \ref{fig:vc}). These discrepancies might be related to the higher progenitor metallicity of SN 2021pfs (Section \ref{sec:hostgalaxy}) or slight differences in explosion physics. Without nebular‑phase spectra for SN 2021pfs, we cannot compare their inner ejecta properties such as nucleosynthetic yields or potential remnant signatures, which are often revealed at $>$+200 days. The lack of late-time spectral and optical data for SN 2021pfs means we should be cautious in declaring it an exact “twin” of SN 2011fe.

We also used the Spectral Adaptive Light curve Template 2 (SALT2, \cite{Guy2005A&A...443..781G,Guy2007A&A...466...11G,Guy2010A&A...523A...7G}) to fit the light curves and derive the parameters and distance moduli for SN 2021pfs, SN 2011fe (\cite{Tsvetkov2013CoSka..43...94T,Graham2015MNRAS.446.2073G,Richmond2012JAVSO..40..872R}), and SN 2011by \citep{Graham2015MNRAS.446.2073G}. The fitting results are presented in Figure \ref{fig:salt_fit} and Table \ref{tab:salt2}. The distance modulus was calculated using the following equation \citep{Guy2010A&A...523A...7G}:
\begin{center}
$\rm DM = \textit{m}_B - \textit{M}_B - \alpha \times \textit{x}_1 -\beta \times \textit{c}$
\end{center}
where $\textit{m}_B, \textit{x}_1, \textit{c}$ are obtained from the SALT2 fit, and $\textit{M}_B, \alpha, \beta $ adopt the values given by 
\cite{Betoule2014A&A...568A..22B} and \cite{Guy2010A&A...523A...7G}.
For SN 2011fe and SN 2011by, the distance moduli derived using the $\textit{M}_B, \alpha, \beta $ parameter set from \cite{Guy2010A&A...523A...7G} are in excellent agreement with their independent Cepheid distances. Including or excluding  \textit{U}-band photometry can noticeably alter the distance modulus derived from the SALT2 fits. This effect is most pronounced for SN 2021pfs: the resulting distance modulus differs by $\sim 0.14$ mag in $\rm DM^{B}$ fits with and without the \textit{U}‑band data, a substantially larger shift than for the other two SNe Ia. In $\mathrm{DM}^{G}$, the corresponding offset is approximately twice as large. In contrast, the effect on the low-redshift SNe 2011fe and 2011by is minimal. As noted by \cite{Dai2023ApJS..267....1D}, including \textit{U}-band data and additional spectra can improve the reliability of SALT2 distances for high-redshift SNe Ia. SN 2021pfs and SN 2011fe have nearly the same light-curve stretch $x_1$ and color $c$ (Table \ref{tab:salt2}). Based on this, the similarity between SN 2021pfs and SN 2011fe is further confirmed. 

\begin{table}
	\centering
	\caption{SALT2 light-curve fit parameters and distance moduli}
	\begin{tabular}{cccccccccc}
		\hline
        \hline
		SN           & Bands     &$ m_{\rm B}$     &$t_0$              &$x_0$          &$x_1$           &$c$                &$\rm DM^{B}$   &$\rm DM^{G}$\\      
		\hline     
		SN 2021pfs   &\textit{BVgri}    &13.74(02)      &59391.960(001)    &0.055(001)    &$-0.283(005)$  &$-0.166(002)$ &33.257(034)  &33.115(059)\\ 
		SN 2021pfs   &\textit{UBVgri}   &13.80(02)      &59391.960(001)    &0.052(001)    &$-0.403(004)$  &$-0.106(001)$ &33.116(033)  &32.833(065)\\ 
 
        SN 2011by    &\textit{BVRI}     &12.98(03)      &55691.072(034)    &0.112(001)    &$-0.249(046)$  &0.067(005)    &31.774(037)  &31.662(082)\\ 
        SN 2011by    &\textit{UBVRI}    &12.95(03)      &55691.180(027)    &0.115(001)    &$-0.076(038)$  &0.038(003)    &31.861(034)  &31.950(067)\\ 
        SN 2011fe    &\textit{BVgr}     &9.98(02)       &55815.447(006)    &1.775(002)    &$-0.304(008)$  &$-0.130(001)$ &29.388(025)  &29.220(055)\\ 
        SN 2011fe    &\textit{UBVgr}    &10.00(02)      &55815.503(006)    &1.746(002)    &$-0.219(007)$  &$-0.118(001)$ &29.380(024)  &29.310(049)\\ 
		\hline
	\end{tabular}
\raggedright Note. \cite{Foley2020MNRAS.491.5991F} used SALT2 to derive distance moduli of DM = 29.16(06) for SN 2011fe and DM = 31.96(04) for SN 2011by. The Cepheid distances to SN 2011by and SN 2011fe are 31.594(071) and 29.135(047) \citep{Foley2020MNRAS.491.5991F,Riess2016ApJ...826...56R}. The superscripts B and G represent the distance moduli calculated using $\textit{M}_B, \alpha, \beta $ from \cite{Betoule2014A&A...568A..22B} and \cite{Guy2010A&A...523A...7G}, respectively.
\label{tab:salt2}
\end{table}

\begin{table}
\centering
\caption{SALT2 light-curve fit parameters and distance moduli}
\begin{tabular}{cccccccccc}
\hline
\hline
SN           & Bands     &$ m_{\rm B}$     &$t_0$              &$x_0$          &$x_1$           &$c$                &$\rm DM^{B}$   &$\rm DM^{G}$\\ 
\hline     
SN 2021pfs   &\textit{BVgri}    &13.79(02)      &59391.960(001)    &0.048(001)    &$-0.222(005)$  &$ 0.052(002)$ &32.633(058)  &32.552(074)\\ 
SN 2021pfs   &\textit{UBVgri}   &13.78(02)      &59391.960(001)    &0.049(001)    &$-0.234(005)$  &$ 0.042(001)$ &32.648(154)  &32.413(170)\\ 
 
SN 2011by    &\textit{BVRI}     &12.98(03)      &55691.054(035)    &0.112(001)    &$-0.219(043)$  &0.065(006)    &31.789(037)  &31.712(078)\\ 
SN 2011by    &\textit{UBVRI}    &12.94(03)      &55691.178(028)    &0.117(001)    &$-0.041(036)$  &0.019(003)    &31.908(033)  &32.038(065)\\ 

SN 2011fe    &\textit{BVgr}     &10.03(02)      &55815.416(006)    &1.556(002)    &$-0.292(007)$  &$-0.062(001)$ &29.218(023)  &29.062(052)\\ 
SN 2011fe    &\textit{UBVgr}    &10.01(02)      &55815.430(006)    &1.578(002)    &$-0.267(007)$  &$-0.051(001)$ &29.175(023)  &29.044(050)\\ 
		\hline
	\end{tabular}
\raggedright Note. \cite{Foley2020MNRAS.491.5991F} used SALT2 to derive distance moduli of DM = 29.16(06) for SN 2011fe and DM = 31.96(04) for SN 2011by. The Cepheid distances to SN 2011by and SN 2011fe are 31.594(071) and 29.135(047) \citep{Foley2020MNRAS.491.5991F,Riess2016ApJ...826...56R}. The superscripts B and G represent the distance moduli calculated using $\textit{M}_B, \alpha, \beta $ from \cite{Betoule2014A&A...568A..22B} and \cite{Guy2010A&A...523A...7G}, respectively.
\label{tab:salt2_new}
\end{table}

\subsection{Host galaxy and Metallicity}\label{sec:hostgalaxy}
The host galaxy of SN 2021pfs, NGC 5427, is an Sc-type spiral galaxy with a Seyfert 2 active nucleus with a stellar mass of $4.61 \times 10^{10} \rm M_{\odot}$ \citep{Weinzirl2009ApJ...696..411W}. To place the progenitor environments of SN~2021pfs and SN~2011fe on a consistent gas-phase metallicity scale, we adopt the O3N2 diagnostic and the calibration of \cite{Pettini2004MNRAS...348L..59P} (hereafter PP04). The O3N2 index is defined as
\begin{equation}
\mathrm{O3N2} \equiv \log\left(\frac{[\rm O{III}]~\lambda5007/\mathrm{H}\beta}{[\rm N{II}]~\lambda6583/\mathrm{H}\alpha}\right),
\end{equation}
and the corresponding oxygen abundance is given by (valid for $-1 < \mathrm{O3N2} < 1.9$)
\begin{equation}
12+\log(\mathrm{O/H}) = 8.73 - 0.32\,\times{\rm O3N2}.
\end{equation}

We estimate the local metallicity of SN~2021pfs using the nearby {H {\sc ii} region reported by \cite{Dopita2014A&A...566A..41D}. From their emission-line ratios for the region closest to the SN site (Table 1, region 36), we adopt
$\log([\rm {O}{III}]/\mathrm{H}\beta)=-0.405$ and
$\log([\rm {N}{II}]/\mathrm{H}\alpha)=-0.450$.
This yields $\mathrm{O3N2} = (-0.405) - (-0.450) = 0.045$, and therefore
\begin{equation}
12+\log(\mathrm{O/H})_{\rm 2021pfs} = 8.73 - 0.32\times 0.045 = 8.717.
\end{equation}

For SN~2011fe, we start from the host-galaxy gas-phase metallicity reported by \citet{Pan2020MNRAS...491.5897P}, which is derived using the PP04 O3N2 calibration. We then estimate the metallicity at the SN position by applying a normalized abundance gradient. Following \citet[][their Eq.~(5)]{Bresolin2007ApJ...656..186B}, we have
\begin{equation}
12+\log(\mathrm{O/H})(R) = \left[12+\log(\mathrm{O/H})\right]_{\rm center} - 0.90\left(\frac{R}{R_0}\right),
\end{equation}
where $R$ is the deprojected galactocentric distance and $R_0=14.4~\arcmin$.
Using the coordinates of SN~2011fe and the galaxy center, we obtain an angular separation of
$\theta \simeq 4.619~\arcmin$.
Given the near face-on geometry of M101 \citep[inclination $i\simeq18^\circ$]{Walter2008AJ....136.2563W}, we calculate than $R \approx \theta/\cos i \simeq 4.857~\arcmin$.
Adopting $\left[12+\log(\mathrm{O/H})\right]_{\rm center}=8.717$ from \cite{Pan2020MNRAS...491.5897P}, we estimate
\begin{equation}
12+\log(\mathrm{O/H})_{\rm 2011fe} \approx 8.717 - 0.90\times\frac{4.857}{14.4} \approx 8.413.
\end{equation}

We emphasize that the abundance gradient of \citet{Bresolin2007ApJ...656..186B} is derived from direct-method (electron-temperature) measurements, whereas the normalization adopted here follows the PP04 O3N2 calibration. As a result, the inferred value at the SN~2011fe position should be regarded as an approximate estimate, and systematic offsets between abundance scales may remain at the level of a few tenths of a dex. Within the above assumptions, the local environment of SN~2021pfs appears more metal-rich than that of SN~2011fe by $\sim0.30$~dex on the adopted scale.

\color{black}It is well established that progenitor metallicity can influence SN Ia luminosity. At a given light‑curve shape, the progenitor metallicity affects the peak luminosity and $^{56}$Ni mass produced in the explosion: higher metallicity implies a fainter supernova and a smaller nickel yield \citep{Timmes2003ApJ...590L..83T,Mazzali2006MNRAS.369L..19M}. This provides a natural explanation for why SN 2021pfs, with its high-metallicity environment, exhibits a lower peak bolometric luminosity and a smaller nickel mass than SN 2011fe. 

Progenitor metallicity may also impact the Si {\sc ii} $\lambda$6355 velocity. \cite{Wang2013Sci...340..170W} and \cite{Pan2015MNRAS.446..354P} found that SNe Ia with higher Si {\sc ii} velocities tend to occur in more massive (hence on average more metal-rich) galaxies and at smaller galactocentric radii. In our case, both SNe 2021pfs and 2011fe belong to the NV and LVG subclasses; however, SN 2011fe shows a slightly higher Si {\sc ii} velocity at maximum than SN 2021pfs, a difference that does not strictly follow the weak metallicity dependence reported for Si {\sc ii} $\lambda$6355 velocity. However, within the NV subclass sample, the range of velocities is small and may not correlate strongly with metallicity \citep {Pan2015MNRAS.446..354P, Burgaz2026A&A...705A..76B}. 
In summary, while SN 2021pfs’s higher host metallicity likely explains its lower Ni mass and luminosity compared to SN 2011fe, it does not appear to have a large effect on the Si {\sc ii} velocity evolution beyond the normal SN Ia diversity.

\begin{figure*}
    \centering
    \includegraphics[width=0.32\linewidth]{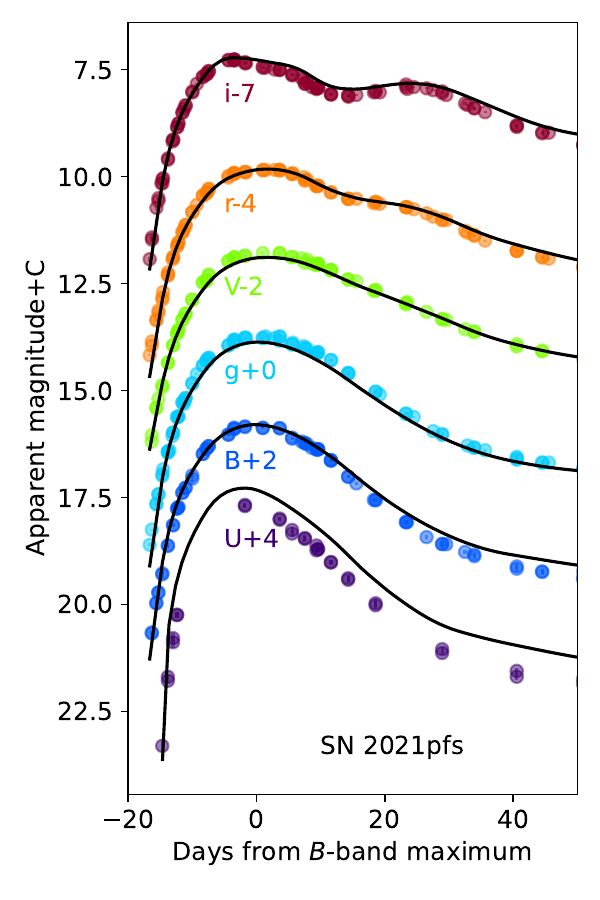}
    \includegraphics[width=0.32\linewidth]{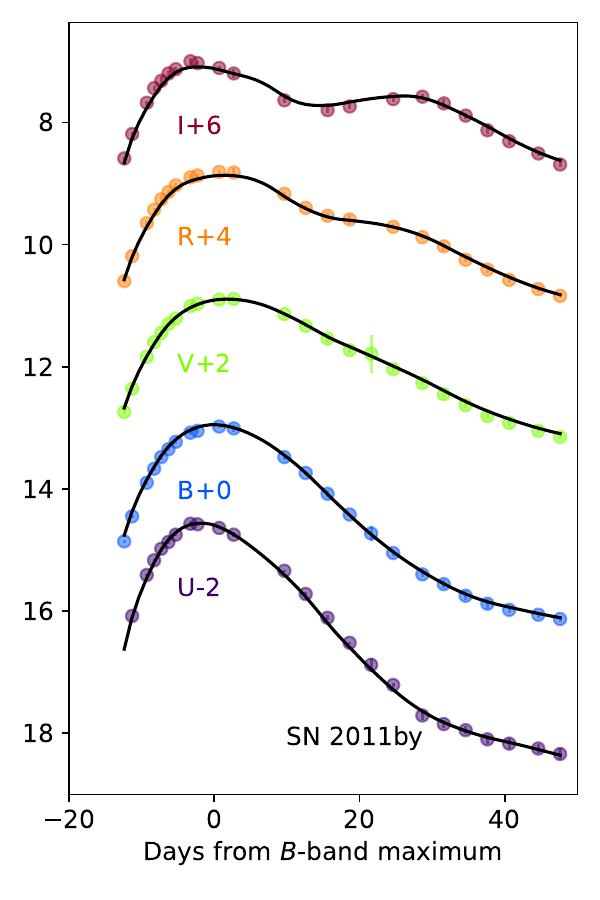}
    \includegraphics[width=0.32\linewidth]{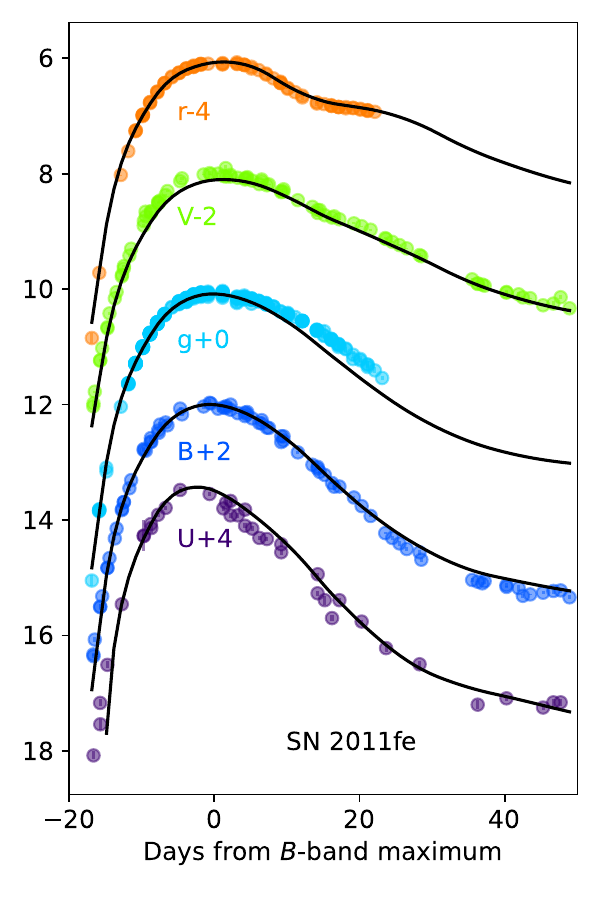}
    \caption{SALT2 light-curves fits for SN 2021pfs, SN 2011fe, and SN 2011by. The observed multi-band light curves (points) are shown with the best-fit SALT2 model fluxes (solid lines) for each SN.}
    \label{fig:salt_fit}
\end{figure*}

\subsection{Explosion Mechanism and Progenitor Constraints}
The presence of unburned carbon and the early light-curve behavior of SN 2021pfs provide clues about its explosion mechanism. In general, a pure carbon deflagration fails to synthesize sufficient $^{56}$Ni ($\textless \, 0.4~M_{\odot}$) to power a normal SN Ia, and instead produces a subluminous event \citep{Barna2021,Liu2023RAA}. Therefore, a transition from deflagration to detonation (delayed detonation, DDT) or a direct carbon detonation  must occur to yield a normal SN Ia. Models that successfully detonate the carbon (either via a DDT near the Chandrasekhar mass, or via a double-detonation in a sub-Chandrasekhar mass WD) are the leading models for explaining normal SNe Ia.

One way to discriminate between models is via nucleosynthesis signatures. The detection of strong C {\sc ii} absorption lines in early spectra indicates that not all the carbon in the outer layers was burned \citep{SN2017cbvHosseinzadeh,iPTF16abc2018,SN2021aefx}. In contrast, sub-Chandrasekhar mass double-detonation and standard Chandrasekhar mass delayed-detonation models tend to burn the high-velocity outer ejecta efficiently and therefore generally predict little or no unburned carbon at the highest velocities \citep{Fink2010A&A...514A..53F,Polin2019ApJ...873...84P}. Despite this tension, a sub-Chandrasekhar mass double-detonation scenario remains attractive for explaining the early photometric behaviour of SN 2021pfs. In spectra earlier than $–$14 days, we observe a broad blend of iron-group elements absorption between $\lambda \approx 4500$--$5000$~\AA\ (Figure \ref{fig:spec}), indicative of enhanced line blanketing that can suppress ultraviolet and blue flux and thus reduce the early-time \textit{U}-band luminosity. 
In the thicker‑shell sub‑Chandrasekhar double-detonation models, radioactive material synthesized in the helium‑shell ashes can produce an early‑time flux excess while simultaneously imprinting strong blended absorption in the UV‑to‑blue spectral region, leading to a significant flux deficit at shorter wavelengths \citep{Polin2019ApJ...873...84P}. 
More generally, mid-UV flux suppression has been attributed to an increased effective UV opacity driven by stronger iron-group line blanketing \citep{DerKacy2023MNRAS}. The early reddening in the \textit{B-V} color evolution further supports this model \citep{Polin2019ApJ...873...84P}. Differences in the early \textit{U}‑and NIR‑band light-curves between the twin candidates SN 2021pfs and SN 2011fe are consistent with predictions from the sub‑Chandrasekhar double‑detonation model. The differences in the early‑time light curves may arise from distinct explosion mechanisms. 

As shown in Figure \ref{fig:C6580}(b), SN 2021pfs clearly shows C {\sc ii} $\lambda$6580 from $-$12 to $-$3 days, suggesting that some unburned carbon is in the outer layers. This is consistent with the LVG SNe Ia \citep{Thomas2011ApJ...743...27T}, though the early $B-V$ color is redder than those of “carbon-positive” SNe Ia that show strong C {\sc ii} (Figure \ref{fig:color}). This combination of weak C II and red early colors aligns with a DDT interpretation: the explosion was essentially a Chandrasekhar-mass DDT, which left a small amount of unburned carbon in the outer ejecta (hence the C {\sc ii} line), and the nickel distribution may have been asymmetric \citep{Polin2021ApJ}, potentially contributing to a slightly lower luminosity (as discussed in Section \ref{sec:Ni}).


\begin{figure}
    \centering
	\includegraphics[width=6in]{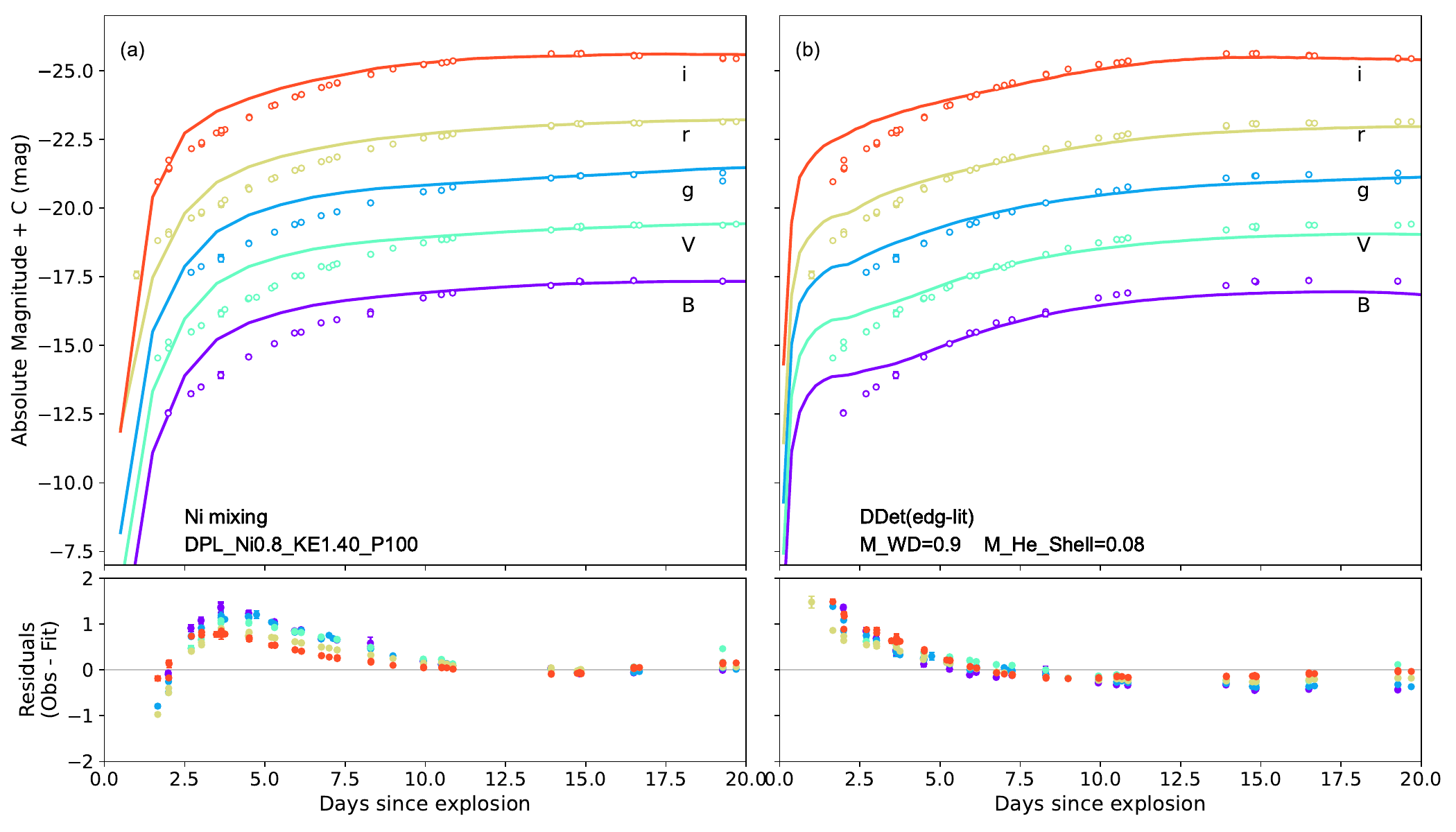}
	\caption{The best matching results of the SN 2021pfs optical light curves and the model mesh, (a): the $^{56}$Ni mixing with the \textbf{$\rm DPL\_Ni0.8\_KE1.40\_P100$} model described by \citep{Magee2020A&A...634A..37M}. (b): the DDet model, the 0.08 M$_{\odot}$ mass of He shell detonation on the 0.9 M$_{\odot}$ mass WD model described by \citep{Polin2019ApJ...873...84P}. 
}
	\label{fig:dde}
\end{figure}

To identify the best-fitting models for the early light curve of SN 2021pfs, we searched grids of both nickel-mixing models \citep{Magee2020A&A...634A..37M} and double-detonation models \citep{Polin2019ApJ...873...84P}. Because no clear excess emission is observed at the earliest phase, we excluded photometry obtained within the first five days after explosion from the fitting. The comparison was then performed using data spanning from day 5 to day 20 after explosion. The best-matching results are shown in Figure \ref{fig:dde}. Within the nickel-mixing grid, the preferred $^{56}$Ni-distribution model is \texttt{DPL\_Ni0.8\_KE1.40\_P100}, which adopts a double-power-law ejecta density profile, a kinetic energy of $1.40\times10^{51}\rm erg$, and 0.8 M$_\odot$ of $^{56}$Ni, with the sharpest transition between the Ni-rich and the Ni-poor zones. This model reproduces the observed light-curve evolution reasonably well between 10 and 20 days after explosion. However, for SN 2021pfs, the $^{56}$Ni mass and kinetic energy inferred from the bolometric light curve (see Section \ref{sec:Ni}) are substantially lower than those required by this model, highlighting a tension between the model fit and the independent bolometric constraints. For the double-detonation grid, the best match is \texttt{good\_0.9\_0.08\_edgelit\_ugrizUBVRI}, corresponding to an edge-lit explosion of a $0.9\,M_{\odot}$ white dwarf with a $0.08\,M_{\odot}$ helium shell \citep{Polin2019ApJ...873...84P}. The model light curves provide a relatively good match to the data from day~5 to day~20 after explosion. Together with the evidence presented above, a sub-Chandrasekhar-mass double-detonation scenario may account for some observed properties of SN~2021pfs. Nevertheless, the lack of an early-time luminosity excess and the presence of only weak carbon absorption are difficult to reconcile with the characteristic predictions of this model.

\section{Conclusion}
\label{sec:con}

Based on a systematic analysis of the observational data for SN 2021pfs, this study confirms it as a spectrally normal SN Ia which is hosted by a Seyfert 2 galaxy with a redshift of $\sim$0.009. The multi-band light curves show that it has a standard peak brightness (absolute magnitude in the \textit{B} band of -19.28\,$\pm$\,0.40\,mag), a standard decay rate ($\Delta m_{15}(B)$\,=\,1.13\,$\pm$\,0.06\,mag), and a standard \textit{B}-band rise time, without significant flux excess in the early stages. Around maximum light spectra show that SN 2021pfs is a member of the Branch-CN group; this conclusion is also consistent with the early ${B-V}$ color. The photospheric velocity measured from Si {\sc ii} $\lambda$6355 line minimum was 10120 $\pm$ 140 km s$^{-1}$, and the velocity gradient was found to be 26.50 $\pm$ 5.00  km s$^{-1} \rm day^{-1}$ after maximum light. Its spectral evolution characteristics are consistent with those of an LVG, NV subclasses of SNe Ia.

A detailed comparison reveals that SN 2021pfs closely resembles SN 2011fe in both light‑curve and spectroscopic evolution, including light‑curve width, absolute‑magnitude light curves, and the evolution of major absorption lines of these spectra. Although both exhibit very similar decline rates, a distinct difference is observed during the rise phase: SN 2021pfs shows a faster rise time than SN 2011fe in shorter wavelengths, whereas it displays a slower rise time in longer wavelengths. SALT2 fitting reveals that such a difference may increase the systematic uncertainty by about 10$\%$ in Type SNe Ia distance estimations. Furthermore, fits to the bolometric light curves indicate that SN 2021pfs has a lower peak luminosity and a smaller nickel yield than SN 2011fe. Further analysis revealed a higher local metallicity at the site of SN 2021pfs compared to SN 2011fe. This metallicity difference may be responsible for the observed discrepancies of these two SNe Ia. SN 2021pfs and SN 2011fe may therefore be considered a potential pair of “twin” SNe Ia. The similarity between them requires further analysis using high-precision distance measurements of NGC 5427. The two objects demonstrate a remarkable resemblance in their overall observational signatures. However, a meticulous comparative analysis identifies non-negligible discrepancies in specific features. Consequently, this study adds weight to the growing body of evidence emphasizing the coexistence of a well-defined uniformity and a significant range of diversity within the observational signatures of SNe Ia.

The sub-Chandrasekhar mass double-detonation model may provide a better explanation for the early-time light curve and spectra features of SN 2021pfs. Spectral analysis reveals the presence of weak C {\sc ii} absorption features in SN 2021pfs between -12 and -3 days, indicating the existence of unburned carbon in the outer layers. This is inconsistent with the sub‑Chandrasekhar‑mass double‑detonation model. Nebula phase spectral data are required to further discuss the theoretical predictions of this model. Furthermore, SN 2021pfs shows no significant early‑time flux excess. A more complex or alternative explosion mechanism may be required to explain the explosion process of this SN.

\section{acknowledgments}
\begin{acknowledgments}
This work was sponsored by the Natural Science Foundation of Xinjiang Uygur Autonomous Region under No. 2024D01D32, the National Natural Science Foundation of China grant 12373038. The Tianshan Talent Training Program, through the grant 2023TSYCCX0101, the Central Guidance for Local Science and Technology Development Fund under No. ZYYD2025QY27, the National Natural Science Foundation of China grants 12433007 and 12541303. The work of X.-F.W. is supported by the National Natural Science Foundation of China (NSFC grants 12288102 and  12033003), the Ma Huateng Foundation, the New Cornerstone Science Foundation through the XPLORER PRIZE, China Manned-Spaced Project (CMS-CSST-2021-A12). W.-X.L. is supported by NSFC (12120101003, 12233008 and 12373010), the National Key R\&D Program (2022YFA1602902 and 2023YFA1607804), Strategic Priority Research Program of CAS (XDB0550100 and XDB0550000), and the Programs of National Astronomical Observatories Chinese Academy of Sciences with Grant Nos. E5ZQ7801, E5ZB7801. This work makes use of observations from the Las Cumbres Observatory network. The LCO team is supported by NSF grants AST-2308113 and AST-1911151. This work was partly supported by the Urumqi Nanshan Astronomy and Deep Space Exploration Observation and Research Station of Xinjiang (XJYWZ2303). NOWT is partly supported by the Operation, Maintenance, and Upgrading Fund for Astronomical Telescopes and Facility Instruments, budgeted from the Ministry of Finance of China (MOF) and administered by the Chinese Academy of Sciences (CAS). L.G. acknowledges financial support from CSIC, MCIN and AEI 10.13039/501100011033 under projects PID2023-151307NB-I00, PIE 20215AT016, and CEX2020-001058-M.
\end{acknowledgments}

\appendix
\setcounter{table}{0}
\renewcommand{\thetable}{A\arabic{table}}
Table \ref{tab_ref} lists the photometric local reference stars in the SN 2021pfs field, and Table \ref{tab:lco_lc} lists the photometry of SN 2021pfs obtained from LCO.

\begin{longrotatetable}
\begin{deluxetable*}{lllrrrrrrll}
\label{tab_ref}
\tablecaption{Photometric Local Reference Stars in the SN 2021pfs Field}
\tablewidth{700pt}
\tabletypesize{\scriptsize}
\tablehead{
\colhead{Star} & 
\colhead{RA(J2000)} & 
\colhead{Dec(J2000)} & 
\colhead{Umag} & 
\colhead{Bmag} & 
\colhead{Vmag} &  
\colhead{gmag} & 
\colhead{rmag} & 
\colhead{imag} \\
} 
\startdata
1  & 210.85927 & -6.05661 & 15.894(026) & 14.865(040) & 13.989(021) & 14.360(023) & 13.718(014) & 13.483(031)\\
2  & 210.81068 & -5.98131 & 18.150(066) & 17.120(004) & 16.494(070) & 16.884(037) & 16.343(054) & 16.155(000)\\
3  & 210.88786 & -6.03869 & 18.312(078) & 17.097(150) & 16.182(070) & 16.598(055) & 15.888(056) & 15.601(149)\\
4  & 210.78537 & -6.06458 & 18.946(053) & 16.874(082) & 15.736(073) & 16.311(035) & 15.252(041) & 14.797(061)\\
5  & 210.86687 & -5.98474 & 16.822(038) & 15.94(057)  & 15.151(031) & 15.490(025) & 14.925(028) & 14.714(050)\\
6  & 210.89291 & -6.09555 & 17.097(072) & 16.451(074) & 15.807(054) & 16.085(042) & 15.641(059) & 15.525(245)\\
7  & 210.89745 & -6.12632 & 15.679(047) & 15.113(045) & 14.454(023) & 14.720(029) & 14.296(038) & 14.088(034)\\
8  & 210.85326 & -6.12268 & 15.999(067) & 15.635(031) & 15.121(023) & 15.320(026) & 15.002(062) & 14.834(115)\\
9  & 210.94036 & -6.01586 & 17.759(098) & 17.341(114) & 16.649(063) & 16.932(021) & 16.558(096) & 16.405(000)\\
10 & 210.87760 & -5.94260 & 17.350(083) & 16.532(092) & 15.814(060) & 16.143(055) & 15.625(063) & 15.418(087)\\
11 & 210.82615 & -5.96049 & 16.323(034) & 15.849(055) & 15.229(051) & 15.475(023) & 15.093(026) & 14.941(094)\\
12 & 210.75708 & -5.99995 & 18.095(048) & 16.623(091) & 15.359(047) & 15.911(026) & 15.023(041) & 14.600(073)\\
13 & 210.76134 & -5.98501 & 18.221(081) & 16.858(122) & 16.003(049) & 16.425(076) & 15.684(029) & 15.397(065)\\
\enddata
\end{deluxetable*}
\end{longrotatetable}

\startlongtable
\begin{deluxetable*}{lcccl}
\tablecolumns{3} 
\tablewidth{5pc} 
\tabletypesize{\scriptsize}
\tablecaption{Photometry of SN 2021pfs taken with LCO}
\tablehead{\colhead{MJD} &\colhead{Epoch$^a$} &\colhead{Mag} &\colhead{Magerr}&\colhead{Filter}}
\startdata
59374.88& 	-16.61&	18.606 	&0.032 	&$g$\\
59374.88& 	-16.61&	18.174 	&0.030 	&$r$\\
59374.88& 	-16.61&	18.932 	&0.057 	&$i$\\
59374.97& 	-16.52&	17.511 	&0.031 	&$z$\\
59374.98& 	-16.51&	17.396 	&0.029 	&$z$\\
59375.19& 	-16.30&	18.195 	&0.031 	&$V$\\
59375.21& 	-16.28&	18.654 	&0.024 	&$B$\\
59375.21& 	-16.28&	18.679 	&0.032 	&$B$\\
59375.21& 	-16.28&	18.047 	&0.047 	&$V$\\
59375.22& 	-16.27&	18.251 	&0.020 	&$g$\\
59375.22& 	-16.27&	17.853 	&0.022 	&$r$\\
59375.22& 	-16.27&	17.949 	&0.021 	&$r$\\
59375.23& 	-16.26&	18.473 	&0.049 	&$i$\\
59375.24& 	-16.25&	18.423 	&0.084 	&$i$\\
59375.74& 	-15.75&	17.718 	&0.098 	&$y$\\
59375.74& 	-15.75&	17.375 	&0.065 	&$y$\\
59375.75& 	-15.74&	16.752 	&0.119 	&$z$\\
59375.75& 	-15.74&	16.843 	&0.122 	&$z$\\
59375.91& 	-15.58&	17.972 	&0.074 	&$B$\\
59375.92& 	-15.57&	17.964 	&0.076 	&$B$\\
59375.92& 	-15.57&	17.405 	&0.057 	&$V$\\
59375.92& 	-15.57&	17.409 	&0.063 	&$V$\\
59375.92& 	-15.57&	17.644 	&0.042 	&$g$\\
59375.93& 	-15.56&	17.661 	&0.041 	&$g$\\
59375.93& 	-15.56&	17.358 	&0.043 	&$r$\\
59375.93& 	-15.56&	17.350 	&0.042 	&$r$\\
59375.93& 	-15.56&	17.730 	&0.049 	&$i$\\
59376.23& 	-15.26&	17.725 	&0.075 	&$B$\\
59376.23& 	-15.26&	17.716 	&0.073 	&$B$\\
59376.24& 	-15.25&	17.196 	&0.060 	&$V$\\
59376.24& 	-15.25&	17.427 	&0.042 	&$g$\\
59376.24& 	-15.25&	17.426 	&0.043 	&$g$\\
59376.24& 	-15.25&	17.186 	&0.037 	&$r$\\
59376.25& 	-15.24&	17.131 	&0.039 	&$r$\\
59376.25& 	-15.24&	17.505 	&0.054 	&$i$\\
59376.25& 	-15.24&	17.567 	&0.055 	&$i$\\
59376.70& 	-14.79&	19.710 	&0.223 	&$u$\\
59376.70& 	-14.79&	17.337 	&0.112 	&$y$\\
59376.70& 	-14.79&	17.001 	&0.093 	&$y$\\
59376.71& 	-14.78&	17.158 	&0.055 	&$i$\\
59376.71& 	-14.78&	16.163 	&0.114 	&$z$\\
59376.71& 	-14.78&	16.131 	&0.107 	&$z$\\
59376.84& 	-14.65&	19.304 	&0.125 	&$U$\\
59376.84& 	-14.65&	17.296 	&0.112 	&$B$\\
59376.85& 	-14.64&	17.273 	&0.107 	&$B$\\
59376.85& 	-14.64&	16.871 	&0.104 	&$V$\\
59376.85& 	-14.64&	16.924 	&0.097 	&$V$\\
59376.85& 	-14.64&	16.951 	&0.092 	&$g$\\
59376.86& 	-14.63&	16.998 	&0.098 	&$g$\\
59376.86& 	-14.63&	16.822 	&0.096 	&$r$\\
59376.86& 	-14.63&	16.876 	&0.096 	&$r$\\
59376.86& 	-14.63&	17.069 	&0.099 	&$i$\\
59376.86& 	-14.63&	17.158 	&0.086 	&$i$\\
59376.95& 	-14.54&	18.992 	&0.536 	&$u$\\
59376.96& 	-14.53&	19.145 	&0.130 	&$u$\\
59376.97& 	-14.52&	16.841 	&0.031 	&$g$\\
59376.97& 	-14.52&	16.697 	&0.034 	&$r$\\
59376.98& 	-14.51&	17.036 	&0.041 	&$i$\\
59377.69& 	-13.80&	16.694 	&0.237 	&$y$\\
59377.69& 	-13.80&	18.333 	&0.155 	&$u$\\
59377.70& 	-13.79&	16.732 	&0.074 	&$y$\\
59377.70& 	-13.79&	15.644 	&0.085 	&$z$\\
59377.70& 	-13.79&	16.457 	&0.049 	&$g$\\
59377.70& 	-13.79&	16.241 	&0.044 	&$r$\\
59377.70& 	-13.79&	15.622 	&0.082 	&$z$\\
59377.71& 	-13.78&	17.691 	&0.044 	&$U$\\
59377.71& 	-13.78&	17.782 	&0.046 	&$U$\\
59377.71& 	-13.78&	16.622 	&0.060 	&$B$\\
59377.72& 	-13.77&	16.630 	&0.058 	&$B$\\
59377.72& 	-13.77&	16.338 	&0.047 	&$V$\\
59377.72& 	-13.77&	16.362 	&0.049 	&$V$\\
59377.72& 	-13.77&	16.432 	&0.037 	&$g$\\
59377.72& 	-13.77&	16.421 	&0.037 	&$g$\\
59377.73& 	-13.76&	16.313 	&0.032 	&$r$\\
59377.73& 	-13.76&	16.301 	&0.031 	&$r$\\
59377.73& 	-13.76&	16.575 	&0.047 	&$i$\\
59377.73& 	-13.76&	16.606 	&0.046 	&$i$\\
59377.96& 	-13.53&	18.121 	&0.683 	&$u$\\
59377.96& 	-13.53&	16.397 	&0.078 	&$g$\\
59378.41& 	-13.08&	17.768 	&0.055 	&$u$\\
59378.42& 	-13.07&	17.711 	&0.054 	&$u$\\
59378.43& 	-13.06&	16.062 	&0.031 	&$g$\\
59378.43& 	-13.06&	15.939 	&0.031 	&$r$\\
59378.43& 	-13.06&	16.182 	&0.039 	&$i$\\
59378.51& 	-12.98&	16.787 	&0.119 	&$U$\\
59378.51& 	-12.98&	16.883 	&0.123 	&$U$\\
59378.51& 	-12.98&	16.151 	&0.043 	&$B$\\
59378.52& 	-12.97&	16.143 	&0.042 	&$B$\\
59378.52& 	-12.97&	15.948 	&0.035 	&$V$\\
59378.52& 	-12.97&	15.941 	&0.034 	&$V$\\
59378.52& 	-12.97&	15.984 	&0.027 	&$g$\\
59378.53& 	-12.96&	15.984 	&0.027 	&$g$\\
59378.53& 	-12.96&	15.885 	&0.029 	&$r$\\
59378.53& 	-12.96&	15.885 	&0.029 	&$r$\\
59378.53& 	-12.96&	16.147 	&0.037 	&$i$\\
59378.53& 	-12.96&	16.135 	&0.035 	&$i$\\
59379.13& 	-12.36&	16.250 	&0.072 	&$U$\\
59379.14& 	-12.35&	16.238 	&0.072 	&$U$\\
59379.14& 	-12.35&	15.761 	&0.030 	&$B$\\
59379.14& 	-12.35&	15.758 	&0.029 	&$B$\\
59379.15& 	-12.34&	15.660 	&0.031 	&$V$\\
59379.15& 	-12.34&	15.672 	&0.031 	&$V$\\
59379.15& 	-12.34&	15.618 	&0.023 	&$g$\\
59379.15& 	-12.34&	15.624 	&0.022 	&$g$\\
59379.16& 	-12.33&	15.620 	&0.024 	&$r$\\
59379.16& 	-12.33&	15.619 	&0.024 	&$r$\\
59379.16& 	-12.33&	15.850 	&0.028 	&$i$\\
59379.16& 	-12.33&	15.842 	&0.028 	&$i$\\
59379.35& 	-12.14&	15.731 	&0.035 	&$B$\\
59379.35& 	-12.14&	15.725 	&0.035 	&$B$\\
59379.35& 	-12.14&	15.588 	&0.030 	&$V$\\
59379.35& 	-12.14&	15.586 	&0.029 	&$V$\\
59379.35& 	-12.14&	15.607 	&0.023 	&$g$\\
59379.36& 	-12.13&	15.607 	&0.024 	&$g$\\
59379.36& 	-12.13&	15.546 	&0.024 	&$r$\\
59379.36& 	-12.13&	15.536 	&0.024 	&$r$\\
59379.36& 	-12.13&	15.760 	&0.034 	&$i$\\
59379.36& 	-12.13&	15.760 	&0.035 	&$i$\\
59379.97& 	-11.52&	15.385 	&0.026 	&$B$\\
59379.97& 	-11.52&	15.392 	&0.026 	&$B$\\
59379.97& 	-11.52&	15.345 	&0.026 	&$V$\\
59379.98& 	-11.51&	15.344 	&0.026 	&$V$\\
59379.98& 	-11.51&	15.278 	&0.019 	&$g$\\
59379.98& 	-11.51&	15.279 	&0.019 	&$g$\\
59379.98& 	-11.51&	15.304 	&0.022 	&$r$\\
59379.98& 	-11.51&	15.301 	&0.022 	&$r$\\
59379.98& 	-11.51&	15.500 	&0.027 	&$i$\\
59379.99& 	-11.50&	15.505 	&0.028 	&$i$\\
59380.22& 	-11.27&	15.314 	&0.027 	&$g$\\
59380.22& 	-11.27&	15.227 	&0.024 	&$r$\\
59380.22& 	-11.27&	15.419 	&0.030 	&$i$\\
59380.34& 	-11.15&	16.462 	&0.063 	&$u$\\
59380.34& 	-11.15&	15.225 	&0.022 	&$g$\\
59380.34& 	-11.15&	15.571 	&0.136 	&$y$\\
59380.35& 	-11.14&	15.594 	&0.142 	&$y$\\
59380.35& 	-11.14&	14.661 	&0.054 	&$z$\\
59380.35& 	-11.14&	14.647 	&0.053 	&$z$\\
59380.46& 	-11.03&	15.273 	&0.031 	&$B$\\
59380.46& 	-11.03&	15.274 	&0.031 	&$B$\\
59380.47& 	-11.02&	15.205 	&0.028 	&$V$\\
59380.47& 	-11.02&	15.204 	&0.027 	&$V$\\
59380.47& 	-11.02&	15.178 	&0.022 	&$g$\\
59380.47& 	-11.02&	15.176 	&0.022 	&$g$\\
59380.47& 	-11.02&	15.134 	&0.023 	&$r$\\
59380.47& 	-11.02&	15.132 	&0.023 	&$r$\\
59380.47& 	-11.02&	15.330 	&0.031 	&$i$\\
59380.48& 	-11.01&	15.353 	&0.032 	&$i$\\
59381.27& 	-10.22&	15.640 	&0.020 	&$y$\\
59381.51& 	-9.98 &15.059 	&0.123 	&$B$\\
59381.51& 	-9.98 &14.983 	&0.039 	&$B$\\
59381.51& 	-9.98 &14.878 	&0.024 	&$V$\\
59381.52& 	-9.97 &14.880 	&0.023 	&$V$\\
59381.52& 	-9.97 &14.834 	&0.019 	&$g$\\
59381.52& 	-9.97 &14.836 	&0.019 	&$g$\\
59381.52& 	-9.97 &14.828 	&0.020 	&$r$\\
59381.52& 	-9.97 &14.834 	&0.020 	&$r$\\
59381.52& 	-9.97 &15.013 	&0.027 	&$i$\\
59381.53& 	-9.96 &15.035 	&0.027 	&$i$\\
59382.21& 	-9.28 &15.698 	&0.043 	&$u$\\
59382.21& 	-9.28 &14.612 	&0.017 	&$g$\\
59382.21& 	-9.28 &14.663 	&0.020 	&$r$\\
59382.21& 	-9.28 &14.833 	&0.026 	&$i$\\
59383.15& 	-8.34 &14.481 	&0.021 	&$B$\\
59383.15& 	-8.34 &14.483 	&0.021 	&$B$\\
59383.16& 	-8.33 &14.473 	&0.020 	&$V$\\
59383.16& 	-8.33 &14.477 	&0.020 	&$V$\\
59383.16& 	-8.33 &14.414 	&0.016 	&$g$\\
59383.16& 	-8.33 &14.416 	&0.016 	&$g$\\
59383.16& 	-8.33 &14.448 	&0.017 	&$r$\\
59383.16& 	-8.33 &14.443 	&0.017 	&$r$\\
59383.17& 	-8.32 &14.677 	&0.022 	&$i$\\
59383.17& 	-8.32 &14.661 	&0.022 	&$i$\\
59383.72& 	-7.77 &14.360 	&0.022 	&$B$\\
59383.72& 	-7.77 &14.362 	&0.023 	&$B$\\
59383.72& 	-7.77 &14.424 	&0.024 	&$V$\\
59383.72& 	-7.77 &14.425 	&0.024 	&$V$\\
59383.73& 	-7.76 &14.303 	&0.019 	&$g$\\
59383.73& 	-7.76 &14.290 	&0.019 	&$g$\\
59383.73& 	-7.76 &14.392 	&0.020 	&$r$\\
59383.73& 	-7.76 &14.381 	&0.020 	&$r$\\
59383.73& 	-7.76 &14.601 	&0.025 	&$i$\\
59383.73& 	-7.76 &14.614 	&0.025 	&$i$\\
59383.88& 	-7.61 &15.319 	&0.099 	&$u$\\
59383.89& 	-7.60 &14.299 	&0.020 	&$g$\\
59383.89& 	-7.60 &14.351 	&0.021 	&$r$\\
59383.89& 	-7.60 &14.586 	&0.026 	&$i$\\
59384.07& 	-7.42 &14.302 	&0.019 	&$B$\\
59384.07& 	-7.42 &14.299 	&0.019 	&$B$\\
59384.07& 	-7.42 &14.296 	&0.018 	&$V$\\
59384.08& 	-7.42 &14.304 	&0.018 	&$V$\\
59384.08& 	-7.41 &14.238 	&0.015 	&$g$\\
59384.08& 	-7.41 &14.237 	&0.015 	&$g$\\
59384.08& 	-7.41 &14.289 	&0.016 	&$r$\\
59384.08& 	-7.41 &14.290 	&0.016 	&$r$\\
59384.08& 	-7.41 &14.539 	&0.022 	&$i$\\
59384.08& 	-7.41 &14.537 	&0.022 	&$i$\\
59387.13& 	-4.36 &14.030 	&0.021 	&$B$\\
59387.13& 	-4.36 &14.031 	&0.021 	&$B$\\
59387.13& 	-4.36 &13.977 	&0.017 	&$V$\\
59387.14& 	-4.35 &13.973 	&0.017 	&$V$\\
59387.14& 	-4.35 &13.948 	&0.015 	&$g$\\
59387.14& 	-4.35 &13.944 	&0.015 	&$g$\\
59387.14& 	-4.35 &14.023 	&0.017 	&$r$\\
59387.14& 	-4.35 &13.984 	&0.015 	&$r$\\
59387.14& 	-4.35 &14.272 	&0.020 	&$i$\\
59387.15& 	-4.34 &14.285 	&0.020 	&$i$\\
59387.95& 	-3.54 &14.852 	&0.078 	&$u$\\
59387.95& 	-3.54 &13.830 	&0.012 	&$g$\\
59387.95& 	-3.54 &13.917 	&0.014 	&$r$\\
59387.95& 	-3.54 &14.274 	&0.019 	&$i$\\
59387.96& 	-3.53 &14.738 	&0.081 	&$y$\\
59387.96& 	-3.53 &13.652 	&0.034 	&$z$\\
59388.02& 	-3.47 &13.867 	&0.014 	&$B$\\
59388.03& 	-3.46 &13.896 	&0.016 	&$V$\\
59388.03& 	-3.46 &14.289 	&0.019 	&$i$\\
59388.06& 	-3.43 &13.897 	&0.016 	&$B$\\
59388.06& 	-3.43 &13.871 	&0.014 	&$g$\\
59388.06& 	-3.43 &13.942 	&0.015 	&$r$\\
59388.06& 	-3.43 &13.896 	&0.016 	&$B$\\
59388.07& 	-3.42 &14.275 	&0.020 	&$i$\\
59388.07& 	-3.42 &13.895 	&0.015 	&$V$\\
59388.07& 	-3.42 &13.889 	&0.015 	&$V$\\
59388.07& 	-3.42 &13.817 	&0.012 	&$g$\\
59388.07& 	-3.42 &13.810 	&0.012 	&$g$\\
59388.07& 	-3.42 &13.916 	&0.014 	&$r$\\
59388.07& 	-3.42 &13.924 	&0.014 	&$r$\\
59388.08& 	-3.41 &14.259 	&0.019 	&$i$\\
59388.08& 	-3.41 &14.272 	&0.019 	&$i$\\
59389.71& 	-1.78 &13.702 	&0.042 	&$U$\\
59389.71& 	-1.78 &13.674 	&0.042 	&$U$\\
59389.71& 	-1.78 &13.853 	&0.019 	&$B$\\
59389.71& 	-1.78 &13.840 	&0.019 	&$B$\\
59389.71& 	-1.78 &13.847 	&0.018 	&$V$\\
59389.71& 	-1.78 &13.855 	&0.019 	&$V$\\
59389.72& 	-1.77 &13.765 	&0.014 	&$g$\\
59389.72& 	-1.77 &13.765 	&0.014 	&$g$\\
59389.72& 	-1.77 &13.890 	&0.017 	&$r$\\
59389.72& 	-1.77 &13.900 	&0.017 	&$r$\\
59389.72& 	-1.77 &14.326 	&0.025 	&$i$\\
59389.72& 	-1.77 &14.356 	&0.026 	&$i$\\
59389.89& 	-1.60 &14.837 	&0.044 	&$u$\\
59389.89& 	-1.60 &13.775 	&0.015 	&$g$\\
59389.89& 	-1.60 &13.903 	&0.017 	&$r$\\
59389.89& 	-1.60 &14.349 	&0.025 	&$i$\\
59392.48& 	0.99 &13.879 	&0.017 	&$B$\\
59392.49& 	1.00 &13.868 	&0.017 	&$B$\\
59392.49& 	1.00 &13.789 	&0.015 	&$V$\\
59392.49& 	1.00 &13.775 	&0.012 	&$g$\\
59392.49& 	1.00 &13.773 	&0.012 	&$g$\\
59392.49& 	1.00 &13.854 	&0.014 	&$r$\\
59392.49& 	1.00 &13.849 	&0.014 	&$r$\\
59392.49& 	1.00 &14.432 	&0.021 	&$i$\\
59392.49& 	1.00 &14.458 	&0.022 	&$i$\\
59392.90& 	1.41 &15.026 	&0.063 	&$u$\\
59392.91& 	1.42 &13.736 	&0.014 	&$g$\\
59392.91& 	1.42 &13.847 	&0.016 	&$r$\\
59392.91& 	1.42 &14.454 	&0.023 	&$i$\\
59394.34& 	2.85 &15.116 	&0.062 	&$u$\\
59394.34& 	2.85 &13.771 	&0.012 	&$g$\\
59394.34& 	2.85 &13.844 	&0.014 	&$r$\\
59394.34& 	2.85 &14.498 	&0.023 	&$i$\\
59394.34& 	2.85 &15.076 	&0.106 	&$y$\\
59394.35& 	2.86 &13.665 	&0.038 	&$z$\\
59395.10& 	3.61 &13.839 	&0.014 	&$g$\\
59395.10& 	3.61 &13.839 	&0.014 	&$r$\\
59395.10& 	3.61 &14.512 	&0.022 	&$i$\\
59395.12& 	3.63 &14.022 	&0.034 	&$U$\\
59395.12& 	3.63 &13.992 	&0.034 	&$U$\\
59395.12& 	3.63 &13.881 	&0.015 	&$B$\\
59395.13& 	3.64 &13.883 	&0.015 	&$B$\\
59395.13& 	3.64 &13.790 	&0.016 	&$V$\\
59395.13& 	3.64 &13.791 	&0.015 	&$V$\\
59395.13& 	3.64 &13.738 	&0.012 	&$g$\\
59395.13& 	3.64 &13.740 	&0.012 	&$g$\\
59395.13& 	3.64 &13.858 	&0.014 	&$r$\\
59395.13& 	3.64 &13.854 	&0.014 	&$r$\\
59395.13& 	3.64 &14.502 	&0.021 	&$i$\\
59395.13& 	3.64 &14.497 	&0.022 	&$i$\\
59397.05& 	5.56 &14.342 	&0.077 	&$U$\\
59397.05& 	5.56 &14.260 	&0.080 	&$U$\\
59397.05& 	5.56 &14.099 	&0.023 	&$B$\\
59397.05& 	5.56 &14.125 	&0.025 	&$B$\\
59397.05& 	5.56 &13.882 	&0.018 	&$V$\\
59397.05& 	5.56 &13.880 	&0.018 	&$V$\\
59397.06& 	5.57 &13.920 	&0.016 	&$g$\\
59397.06& 	5.57 &13.925 	&0.017 	&$g$\\
59397.06& 	5.57 &13.938 	&0.016 	&$r$\\
59397.06& 	5.57 &13.946 	&0.016 	&$r$\\
59397.06& 	5.57 &14.635 	&0.026 	&$i$\\
59397.06& 	5.57 &14.618 	&0.024 	&$i$\\
59397.16& 	5.67 &15.425 	&0.048 	&$u$\\
59397.16& 	5.67 &13.882 	&0.013 	&$g$\\
59397.16& 	5.67 &13.917 	&0.014 	&$r$\\
59397.16& 	5.67 &14.613 	&0.022 	&$i$\\
59398.53& 	7.04 &14.192 	&0.020 	&$B$\\
59398.53& 	7.04 &13.912 	&0.016 	&$V$\\
59398.53& 	7.04 &14.702 	&0.024 	&$i$\\
59398.88& 	7.39 &15.615 	&0.050 	&$u$\\
59398.88& 	7.39 &13.965 	&0.013 	&$g$\\
59398.88& 	7.39 &14.016 	&0.015 	&$r$\\
59398.88& 	7.39 &14.737 	&0.025 	&$i$\\
59398.88& 	7.39 &15.346 	&0.131 	&$y$\\
59398.89& 	7.40 &13.849 	&0.040 	&$z$\\
59399.04& 	7.55 &14.468 	&0.074 	&$U$\\
59399.04& 	7.55 &14.447 	&0.076 	&$U$\\
59399.04& 	7.55 &14.231 	&0.028 	&$B$\\
59399.05& 	7.56 &14.227 	&0.029 	&$B$\\
59399.05& 	7.56 &14.017 	&0.037 	&$V$\\
59399.05& 	7.56 &14.020 	&0.022 	&$g$\\
59399.05& 	7.56 &14.016 	&0.021 	&$g$\\
59399.05& 	7.56 &14.089 	&0.024 	&$r$\\
59399.05& 	7.56 &14.090 	&0.027 	&$r$\\
59399.05& 	7.56 &14.834 	&0.047 	&$i$\\
59399.05& 	7.56 &14.827 	&0.042 	&$i$\\
59399.40& 	7.91 &15.679 	&0.058 	&$u$\\
59399.40& 	7.91 &14.022 	&0.015 	&$g$\\
59399.40& 	7.91 &14.068 	&0.016 	&$r$\\
59399.40& 	7.91 &14.792 	&0.026 	&$i$\\
59399.42& 	7.93 &14.258 	&0.020 	&$B$\\
59399.42& 	7.93 &13.945 	&0.016 	&$V$\\
59399.42& 	7.93 &14.793 	&0.025 	&$i$\\
59400.13& 	8.64 &14.341 	&0.028 	&$B$\\
59400.13& 	8.64 &14.032 	&0.023 	&$V$\\
59400.13& 	8.64 &14.911 	&0.045 	&$i$\\
59400.85& 	9.36 &14.735 	&0.053 	&$U$\\
59400.85& 	9.36 &14.622 	&0.054 	&$U$\\
59400.85& 	9.36 &14.373 	&0.023 	&$B$\\
59400.85& 	9.36 &14.380 	&0.023 	&$B$\\
59400.86& 	9.37 &14.067 	&0.020 	&$V$\\
59400.86& 	9.37 &14.070 	&0.020 	&$V$\\
59400.86& 	9.37 &14.126 	&0.017 	&$g$\\
59400.86& 	9.37 &14.124 	&0.017 	&$g$\\
59400.86& 	9.37 &14.222 	&0.018 	&$r$\\
59400.86& 	9.37 &14.221 	&0.018 	&$r$\\
59400.86& 	9.37 &14.929 	&0.029 	&$i$\\
59400.86& 	9.37 &14.958 	&0.029 	&$i$\\
59400.87& 	9.38 &15.820 	&0.065 	&$u$\\
59400.87& 	9.38 &14.141 	&0.018 	&$g$\\
59400.87& 	9.38 &14.216 	&0.020 	&$r$\\
59400.87& 	9.38 &14.934 	&0.032 	&$i$\\
59401.09& 	9.60 &14.712 	&0.044 	&$U$\\
59401.09& 	9.60 &14.701 	&0.044 	&$U$\\
59401.09& 	9.60 &14.378 	&0.018 	&$B$\\
59401.09& 	9.60 &14.369 	&0.018 	&$B$\\
59401.09& 	9.60 &14.037 	&0.017 	&$V$\\
59401.10& 	9.61 &14.042 	&0.017 	&$V$\\
59401.10& 	9.61 &14.105 	&0.013 	&$g$\\
59401.10& 	9.61 &14.108 	&0.013 	&$g$\\
59401.10& 	9.61 &14.220 	&0.016 	&$r$\\
59401.10& 	9.61 &14.210 	&0.016 	&$r$\\
59401.10& 	9.61 &14.930 	&0.025 	&$i$\\
59401.10& 	9.61 &14.907 	&0.025 	&$i$\\
59401.11& 	9.62 &14.377 	&0.018 	&$B$\\
59401.11& 	9.62 &14.040 	&0.017 	&$V$\\
59401.11& 	9.62 &14.921 	&0.025 	&$i$\\
59403.10& 	11.61 &15.020 	&0.050 	&$U$\\
59403.10& 	11.61 &15.009 	&0.051 	&$U$\\
59403.10& 	11.61 &14.627 	&0.021 	&$B$\\
59403.10& 	11.61 &14.631 	&0.020 	&$B$\\
59403.11& 	11.62 &14.188 	&0.018 	&$V$\\
59403.11& 	11.62 &14.183 	&0.018 	&$V$\\
59403.11& 	11.62 &14.293 	&0.014 	&$g$\\
59403.11& 	11.62 &14.287 	&0.014 	&$g$\\
59403.11& 	11.62 &14.373 	&0.017 	&$r$\\
59403.11& 	11.62 &14.367 	&0.017 	&$r$\\
59403.11& 	11.62 &15.076 	&0.027 	&$i$\\
59403.11& 	11.62 &15.067 	&0.027 	&$i$\\
59403.12& 	11.63 &14.640 	&0.023 	&$B$\\
59403.12& 	11.63 &14.189 	&0.018 	&$V$\\
59403.12& 	11.63 &15.087 	&0.027 	&$i$\\
59405.78& 	14.29 &15.422 	&0.085 	&$U$\\
59405.78& 	14.29 &15.388 	&0.084 	&$U$\\
59405.78& 	14.29 &15.013 	&0.031 	&$B$\\
59405.78& 	14.29 &15.014 	&0.031 	&$B$\\
59405.78& 	14.29 &14.409 	&0.024 	&$V$\\
59405.78& 	14.29 &14.401 	&0.024 	&$V$\\
59405.78& 	14.29 &14.593 	&0.021 	&$g$\\
59405.79& 	14.30 &14.589 	&0.021 	&$g$\\
59405.79& 	14.30 &14.520 	&0.022 	&$r$\\
59405.79& 	14.30 &14.522 	&0.022 	&$r$\\
59405.79& 	14.30 &15.107 	&0.034 	&$i$\\
59405.79& 	14.30 &15.106 	&0.035 	&$i$\\
59405.83& 	14.34 &16.614 	&0.036 	&$u$\\
59405.83& 	14.34 &14.610 	&0.021 	&$g$\\
59405.83& 	14.34 &14.519 	&0.021 	&$r$\\
59405.83& 	14.34 &15.135 	&0.031 	&$i$\\
59405.83& 	14.34 &15.322 	&0.126 	&$y$\\
59405.84& 	14.35 &14.004 	&0.044 	&$z$\\
59407.06& 	15.57 &15.172 	&0.029 	&$B$\\
59407.06& 	15.57 &14.450 	&0.020 	&$V$\\
59407.06& 	15.57 &15.093 	&0.026 	&$i$\\
59407.06& 	15.57 &14.520 	&0.017 	&$r$\\
59409.78& 	18.29 &15.558 	&0.040 	&$B$\\
59409.78& 	18.29 &14.677 	&0.027 	&$V$\\
59409.78& 	18.29 &15.033 	&0.034 	&$i$\\
59409.78& 	18.29 &14.625 	&0.023 	&$r$\\
59410.05& 	18.56 &15.973 	&0.091 	&$U$\\
59410.06& 	18.57 &16.019 	&0.091 	&$U$\\
59410.06& 	18.57 &15.567 	&0.032 	&$B$\\
59410.06& 	18.57 &15.558 	&0.033 	&$B$\\
59410.06& 	18.57 &14.657 	&0.021 	&$V$\\
59410.06& 	18.57 &14.653 	&0.021 	&$V$\\
59410.06& 	18.57 &15.025 	&0.020 	&$g$\\
59410.06& 	18.57 &15.033 	&0.020 	&$g$\\
59410.06& 	18.57 &14.594 	&0.018 	&$r$\\
59410.06& 	18.57 &14.594 	&0.019 	&$r$\\
59410.06& 	18.57 &15.021 	&0.027 	&$i$\\
59410.07& 	18.58 &14.997 	&0.027 	&$i$\\
59410.70& 	19.21 &17.228 	&0.102 	&$u$\\
59410.70& 	19.21 &15.092 	&0.025 	&$g$\\
59410.70& 	19.21 &14.637 	&0.022 	&$r$\\
59410.70& 	19.21 &15.034 	&0.030 	&$i$\\
59410.70& 	19.21 &15.210 	&0.123 	&$y$\\
59410.71& 	19.22 &13.951 	&0.044 	&$z$\\
59414.78& 	23.29 &16.070 	&0.054 	&$B$\\
59414.78& 	23.29 &16.081 	&0.053 	&$B$\\
59414.78& 	23.29 &14.990 	&0.031 	&$V$\\
59414.78& 	23.29 &14.960 	&0.030 	&$V$\\
59414.78& 	23.29 &15.540 	&0.032 	&$g$\\
59414.78& 	23.29 &15.542 	&0.032 	&$g$\\
59414.79& 	23.30 &14.714 	&0.024 	&$r$\\
59414.79& 	23.30 &14.718 	&0.024 	&$r$\\
59414.79& 	23.30 &14.908 	&0.032 	&$i$\\
59414.79& 	23.30 &14.849 	&0.032 	&$i$\\
59414.97& 	23.48 &16.075 	&0.039 	&$B$\\
59414.97& 	23.48 &14.935 	&0.023 	&$V$\\
59414.97& 	23.48 &14.951 	&0.025 	&$i$\\
59414.97& 	23.48 &14.719 	&0.019 	&$r$\\
59415.99& 	24.50 &17.793 	&0.095 	&$u$\\
59415.99& 	24.50 &15.616 	&0.024 	&$g$\\
59415.99& 	24.50 &14.765 	&0.019 	&$r$\\
59415.99& 	24.50 &14.909 	&0.024 	&$i$\\
59415.99& 	24.50 &14.834 	&0.085 	&$y$\\
59416.00& 	24.51 &13.988 	&0.037 	&$z$\\
59417.98& 	26.49 &16.426 	&0.054 	&$B$\\
59417.99& 	26.50 &15.162 	&0.027 	&$V$\\
59417.99& 	26.50 &14.938 	&0.025 	&$i$\\
59417.99& 	26.50 &14.859 	&0.021 	&$r$\\
59419.02& 	27.53 &18.036 	&0.112 	&$u$\\
59419.03& 	27.54 &15.949 	&0.027 	&$g$\\
59419.03& 	27.54 &14.935 	&0.022 	&$r$\\
59419.03& 	27.54 &14.981 	&0.026 	&$i$\\
59420.42& 	28.93 &17.046 	&0.051 	&$U$\\
59420.42& 	28.93 &17.132 	&0.056 	&$U$\\
59420.42& 	28.93 &16.590 	&0.056 	&$B$\\
59420.42& 	28.93 &16.585 	&0.055 	&$B$\\
59420.42& 	28.93 &15.335 	&0.031 	&$V$\\
59420.42& 	28.93 &15.329 	&0.031 	&$V$\\
59420.42& 	28.93 &16.014 	&0.034 	&$g$\\
59420.43& 	28.94 &16.039 	&0.035 	&$g$\\
59420.43& 	28.94 &15.019 	&0.024 	&$r$\\
59420.43& 	28.94 &15.002 	&0.024 	&$r$\\
59420.43& 	28.94 &15.049 	&0.030 	&$i$\\
59420.43& 	28.94 &15.011 	&0.030 	&$i$\\
59421.05& 	29.56 &16.593 	&0.050 	&$B$\\
59421.05& 	29.56 &15.354 	&0.028 	&$V$\\
59421.05& 	29.56 &15.095 	&0.027 	&$i$\\
59421.05& 	29.56 &15.018 	&0.022 	&$r$\\
59423.98& 	32.49 &16.772 	&0.041 	&$B$\\
59423.98& 	32.49 &15.569 	&0.029 	&$V$\\
59423.98& 	32.49 &15.291 	&0.027 	&$i$\\
59423.98& 	32.49 &15.268 	&0.024 	&$r$\\
59424.35& 	32.86 &18.219 	&0.083 	&$u$\\
59424.36& 	32.87 &16.285 	&0.036 	&$g$\\
59424.36& 	32.87 &15.311 	&0.026 	&$r$\\
59424.36& 	32.87 &15.327 	&0.032 	&$i$\\
59425.42& 	33.93 &16.834 	&0.053 	&$B$\\
59425.42& 	33.93 &16.873 	&0.053 	&$B$\\
59425.42& 	33.93 &15.615 	&0.032 	&$V$\\
59425.42& 	33.93 &15.639 	&0.032 	&$V$\\
59425.42& 	33.93 &16.332 	&0.031 	&$g$\\
59425.42& 	33.93 &16.340 	&0.032 	&$g$\\
59425.43& 	33.94 &15.364 	&0.025 	&$r$\\
59425.43& 	33.94 &15.387 	&0.026 	&$r$\\
59425.43& 	33.94 &15.406 	&0.032 	&$i$\\
59425.43& 	33.94 &15.401 	&0.032 	&$i$\\
59427.09& 	35.60 &18.367 	&0.220 	&$u$\\
59427.10& 	35.61 &16.397 	&0.031 	&$g$\\
59427.10& 	35.61 &15.497 	&0.026 	&$r$\\
59427.10& 	35.61 &15.496 	&0.030 	&$i$\\
59432.03& 	40.54 &17.552 	&0.076 	&$U$\\
59432.03& 	40.54 &17.688 	&0.086 	&$U$\\
59432.03& 	40.54 &17.110 	&0.072 	&$B$\\
59432.03& 	40.54 &17.163 	&0.074 	&$B$\\
59432.03& 	40.54 &18.578 	&0.176 	&$u$\\
59432.03& 	40.54 &15.918 	&0.038 	&$V$\\
59432.03& 	40.54 &15.973 	&0.039 	&$V$\\
59432.03& 	40.54 &16.618 	&0.036 	&$g$\\
59432.03& 	40.54 &16.550 	&0.039 	&$g$\\
59432.03& 	40.54 &15.733 	&0.028 	&$r$\\
59432.03& 	40.54 &16.593 	&0.041 	&$g$\\
59432.04& 	40.55 &15.842 	&0.036 	&$i$\\
59432.04& 	40.55 &15.742 	&0.030 	&$r$\\
59432.04& 	40.55 &15.731 	&0.030 	&$r$\\
59432.04& 	40.55 &15.800 	&0.036 	&$i$\\
59432.04& 	40.55 &15.815 	&0.036 	&$i$\\
59435.98& 	44.49 &17.228 	&0.066 	&$B$\\
59435.99& 	44.50 &17.240 	&0.065 	&$B$\\
59435.99& 	44.50 &16.083 	&0.037 	&$V$\\
59435.99& 	44.50 &16.071 	&0.037 	&$V$\\
59435.99& 	44.50 &16.687 	&0.036 	&$g$\\
59435.99& 	44.50 &16.684 	&0.036 	&$g$\\
59435.99& 	44.50 &15.892 	&0.029 	&$r$\\
59436.00& 	44.51 &15.876 	&0.028 	&$r$\\
59436.00& 	44.51 &15.967 	&0.034 	&$i$\\
59436.00& 	44.51 &15.990 	&0.035 	&$i$\\
59437.02& 	45.53 &18.845 	&0.494 	&$u$\\
59437.02& 	45.53 &16.683 	&0.039 	&$g$\\
59437.02& 	45.53 &15.920 	&0.029 	&$r$\\
59437.02& 	45.53 &15.975 	&0.035 	&$i$\\
59442.36& 	50.87 &17.855 	&0.066 	&$U$\\
59442.36& 	50.87 &17.414 	&0.073 	&$B$\\
59442.38& 	50.89 &17.743 	&0.057 	&$U$\\
59442.38& 	50.89 &17.231 	&0.070 	&$B$\\
59442.38& 	50.89 &17.362 	&0.070 	&$B$\\
59442.39& 	50.90 &16.255 	&0.043 	&$V$\\
59442.39& 	50.90 &16.260 	&0.043 	&$V$\\
59442.39& 	50.90 &16.807 	&0.043 	&$g$\\
59442.39& 	50.90 &19.234 	&0.234 	&$u$\\
59442.39& 	50.90 &16.843 	&0.042 	&$g$\\
59442.39& 	50.90 &16.859 	&0.042 	&$g$\\
59442.39& 	50.90 &16.094 	&0.035 	&$r$\\
59442.39& 	50.90 &16.145 	&0.036 	&$r$\\
59442.39& 	50.90 &16.279 	&0.046 	&$i$\\
59442.39& 	50.90 &16.115 	&0.035 	&$r$\\
59442.40& 	50.91 &16.284 	&0.045 	&$i$\\
59442.40& 	50.91 &16.238 	&0.045 	&$i$\\
59450.73& 	59.24 &16.532 	&0.054 	&$V$\\
59450.73& 	59.24 &16.541 	&0.054 	&$V$\\
59450.73& 	59.24 &16.984 	&0.046 	&$g$\\
59450.73& 	59.24 &16.971 	&0.046 	&$g$\\
59450.74& 	59.25 &16.411 	&0.039 	&$r$\\
59450.74& 	59.25 &16.421 	&0.040 	&$r$\\
59450.74& 	59.25 &16.574 	&0.046 	&$i$\\
59450.74& 	59.25 &16.580 	&0.046 	&$i$\\
59451.72& 	60.23 &16.959 	&0.066 	&$g$\\
59456.01& 	64.52 &17.587 	&0.057 	&$B$\\
59456.01& 	64.52 &17.552 	&0.057 	&$B$\\
59456.01& 	64.52 &16.602 	&0.041 	&$V$\\
59456.01& 	64.52 &16.636 	&0.041 	&$V$\\
59456.01& 	64.52 &17.071 	&0.038 	&$g$\\
59456.02& 	64.53 &17.076 	&0.038 	&$g$\\
59456.02& 	64.53 &16.597 	&0.035 	&$r$\\
59456.02& 	64.53 &16.604 	&0.035 	&$r$\\
59456.02& 	64.53 &16.797 	&0.041 	&$i$\\
59456.02& 	64.53 &16.793 	&0.045 	&$i$\\
59458.02& 	66.53 &16.678 	&0.027 	&$r$\\
59458.02& 	66.53 &16.896 	&0.026 	&$i$\\
59463.98& 	72.49 &17.797 	&0.027 	&$B$\\
59463.98& 	72.49 &17.703 	&0.024 	&$B$\\
59463.99& 	72.50 &16.960 	&0.019 	&$V$\\
59463.99& 	72.50 &16.861 	&0.019 	&$V$\\
59464.38& 	72.89 &17.240 	&0.044 	&$g$\\
59464.38& 	72.89 &17.209 	&0.044 	&$g$\\
59464.39& 	72.90 &16.868 	&0.015 	&$r$\\
59464.39& 	72.90 &16.864 	&0.016 	&$r$\\
59464.39& 	72.90 &17.037 	&0.029 	&$i$\\
59464.39& 	72.90 &17.212 	&0.032 	&$i$\\
59470.98& 	79.49 &17.367 	&0.053 	&$g$\\
59470.98& 	79.49 &17.150 	&0.035 	&$r$\\
59470.98& 	79.49 &17.250 	&0.067 	&$i$\\
\enddata
\tablenotetext{a}{Relative to the epoch of $B$-band maximum brightness (MJD = 59391.49).}
\label{tab:lco_lc}
\end{deluxetable*}

\bibliography{PASPsample631}{}
\bibliographystyle{aasjournal}



\end{document}